\documentclass[12pt]{report}

\usepackage[dvips]{psfrag}

\usepackage{amsfonts}
\usepackage{amsmath}
\usepackage{amssymb}
\usepackage{bm}
\usepackage{graphicx}
\usepackage{epsf}
\usepackage[T1]{fontenc}
\usepackage{a4wide}
\usepackage{setspace}
\usepackage{cite}
\usepackage[T1]{fontenc}

\author{Yaron Bromberg}
 

\begin{document}
\pagenumbering{roman}
\begin{singlespace}
\begin{center}\thispagestyle{empty}\includegraphics{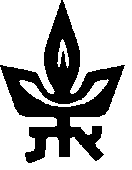}{\large \vspace{0cm}}\end{center}{\large \par}
\thispagestyle{empty}
\end{singlespace}

\begin{center}{\large Tel-Aviv University}\end{center}{\large \par}

\begin{center}{\large Raymond and Beverly Sackler}\end{center}{\large \par}

\begin{center}{\large Faculty of Exact Sciences}\end{center}{\large \par}

\begin{center}{\large School of Physics and Astronomy} {{\large \vspace{1.0cm}}}\end{center}{\large \par}

\begin{center}\textbf{\LARGE Response of Discrete Nonlinear }\end{center}{\LARGE \par}

\begin{center}\textbf{\LARGE Systems With Many Degrees of Freedom} {\textbf{\LARGE \vfill}}\end{center}{\LARGE \par}

\begin{doublespace}
\begin{center}{\large Thesis submitted by} \textbf{\large Yaron Bromberg}
{\large in partial fulfillment of the requirements for a M.Sc.
degree in physics.}\end{center}{\large \par}
\end{doublespace}

\begin{center}{\large This work was carried out under the supervision
of }\end{center}{\large \par}

\begin{center}\textbf{\large Dr. Ron Lifshitz}{\large .} {{\large \vspace{1.2cm}}}\end{center}{\large \par}

\begin{center}\textbf{\large October 2004}\end{center}{\large \par}
\begin{center}{revised on October 2005}\end{center}{\large \par}
\doublespacing
\newpage
\begin{abstract} We study the response of a large array of coupled
nonlinear oscillators to parametric excitation, motivated by the
growing interest in the nonlinear dynamics of microelectromechanical
and nanoelectromechanical systems (MEMS and NEMS). Using a
multiscale analysis, we derive an amplitude equation that captures
the slow dynamics of the coupled oscillators just above the onset of
parametric oscillations. The amplitude equation that we derive here
from first principles contains uncommon nonlinear gradient terms
which yield a unique wave-number dependent bifurcation similar in
character to the behavior known to exist in fluids undergoing the
Faraday wave instability. We suggest a number of experiments with
nanomechanical or micromechanical resonators to test the predictions
of our theory, in particular the strong hysteretic dependence on the
drive amplitude.
\end{abstract}

\newpage
\tableofcontents
\newpage
\section*{Acknowledgements} I would like to express my deepest
gratitude to my supervisor Dr. Ron Lifhsitz, who introduced me to
independent research work, yet helped me focus on my work with
invaluable comments and suggestions. He was always available, not
only for questions and discussions, but also for assistance with
every problem or difficulty that came up. I am particularly grateful
for the many hours he spent working with me on the presentation of
our results.

It was a unique privilege to learn the basic concepts of the theory
of amplitude equations from Prof. Michael Cross from The California
Institute of Technology. His clear and patient answers to my endless
questions while working with him at the Physbio workshop in
Benasque, Spain, and during the past year, were of utmost value to
this work.

I wish to thank my fellow students at the physics school for many
fruitful discussions, and a special thanks to my friend Etay Mar-Or
for his invaluable assistance with the numerical work presented
here.

I thank the U.S.-Israel Binational Foundation (BSF) for their
support of this research under Grant No.~1999458.

Finally, it is a pleasure to thank my family who supported me all
the way and always had a good advice for me.

\newpage
\setcounter{page}{1} \pagenumbering{arabic}
\addcontentsline{toc}{chapter}{Introduction}
\chapter*{Introduction}
In the last decade we have witnessed exciting technological advances
in the fabrication and control of microelectromechanical and
nanoelectromechanical systems (MEMS and NEMS). Such systems are
being developed for a host of nanotechnological applications, as
well as for basic research in the mesoscopic physics of phonons, and
the general study of the behavior of mechanical degrees of freedom
at the interface between the quantum and the classical
worlds~\cite{R01,C03,blencowe}.  Surprisingly, NEMS have also opened
up a new experimental window into the study of the nonlinear
dynamics of discrete systems with many degrees of freedom. A
combination of three properties of NEMS resonators has led to this
unique experimental opportunity. First and most important is the
experimental observation that micro- and nanomechanical resonators
tend to behave nonlinearly at very modest amplitudes. This nonlinear
behavior has not only been observed
experimentally~\cite{turner98,C00,BR01,blick02,turner02,turner03,yu02,cleland04},
but has already been exploited to achieve mechanical signal
amplification and mechanical noise squeezing~\cite{rugar,carr} in
single resonators.  Second is the fact that at their dimensions, the
normal frequencies of nanomechanical resonators are extremely
high---recently exceeding the 1GHz mark~\cite{HZMR03}---facilitating
the design of ultra-fast mechanical devices, and making the waiting
times for unwanted transients bearable on experimental time scales.
Third is the technological ability to fabricate large arrays of MEMS
and NEMS resonators whose collective response might be useful for
signal enhancement and noise reduction, as well as for sophisticated
mechanical signal processing applications. Such arrays have already
exhibited interesting nonlinear dynamics ranging from the formation
of extended patterns~\cite{BR02}---as one commonly observes in
analogous continuous systems such as Faraday waves---to that of
intrinsically localized modes~\cite{SHSICC03}. Thus, nanomechanical
resonator arrays are perfect for testing dynamical theories of
discrete nonlinear systems with many degrees of freedom.  At the
same time, the theoretical understanding of such systems may prove
useful for future nanotechnological applications.

This work is motivated by a recent experiment of Buks and
Roukes~\cite{BR02}, who succeeded in fabricating, exciting, and
measuring the response to parametric excitation of an array of 67
micromechanical resonating gold beams. Lifshitz and
Cross~\cite{LC03} described the response of the beams with a set of
coupled nonlinear differential equations~(\ref{eom}). They used
secular perturbation theory to convert these equations into a set of
coupled nonlinear {\it algebraic\/} equations for the normal mode
amplitudes of the system, enabling them to obtain exact results for
small arrays but only a qualitative understanding of the dynamics of
large arrays. In order to obtain analytical results for large arrays
we study here the same system of equations, approaching it from the
continuous limit of infinitely-many degrees of freedom. Our central
result is a scaled amplitude equation~(\ref{BampEq}), governed by a
single control parameter, that captures the slow dynamics of the
coupled oscillators just above the onset of parametric oscillations.
This amplitude equation includes uncommon nonlinear gradient terms
and exhibits a unique wave-number dependent bifurcation that can be
traced back to the parameters of the equations of motion of the
system~(\ref{eom}). We confirm this behavior numerically and make
suggestions for testing it experimentally.

This thesis is organized as follows. Chapters~1-3 serve as a brief
background, summarizing the results of Buks and Roukes (chapter~1)
and Lifshitz and Cross (chapters~2,3) that are essential for the
understanding of the work presented in the following chapters. In
chapter~4 we present a detailed description of our derivation of the
amplitude equations describing the response of large arrays. A
reduction to a single amplitude equation just above the onset of
oscillations is preformed in chapter~5. Single mode solutions of the
amplitude equation are discussed in chapter~6, and experimental
schemes for obtaining such solutions and their unique properties are
suggested and demonstrated numerically in chapter~7.

\chapter{Experimental Motivation}\label{BRexperiment}
\begin{figure}
\begin{center}
\includegraphics[width=0.6\textwidth]{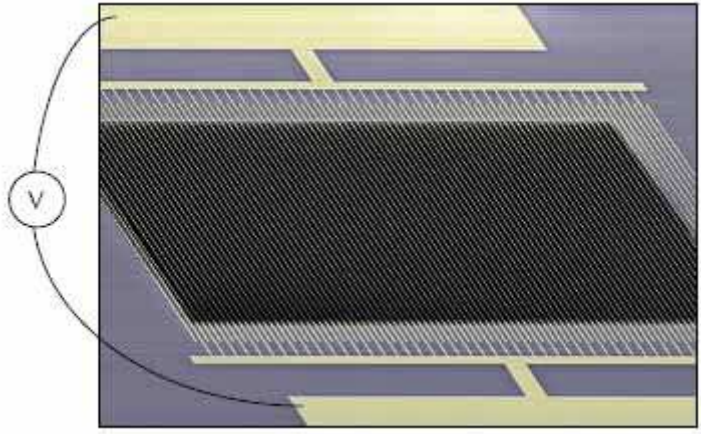}%
\caption{\label{fig:deivce} A side view micrograph of the array
constructed by Buks and Roukes. The purple background is the silicon
surface, from which a rectangular piece was etched away (black
rectangular), leaving the fabricated gold beams connected to the
silicon surface from both edges. A potential difference $V$ is
applied between the two base electrodes.}
\end{center}
\end{figure}

This work is motivated by a recent experiment of Buks and Roukes
~\cite[henceforth BR]{BR02} who constructed an array of
micromechanical resonating beams forming a diffraction grating.
Fig.~\ref{fig:deivce} shows a micrograph of 67 gold beams that BR
fabricated on top of a silicon nitride membrane. Each beam was
$270\times 1\times0.25$ $\mu m$ in size, and the distance between
two neighboring beams was $4\mu m$. The membrane was then removed,
leaving only the ends of the beams connected to the silicon surface.
Electrostatic forces between the beams were applied by connecting
them alternately to two base electrodes. Such forces induce tunable
coupling between the beams, and thus a collective spectrum of
vibrational modes could be obtained. By introducing an ac component
to the electrostatic forces, these modes can be parametrically
excited. Optical diffraction was used to study the response of the
array as a function of the driving frequency and the dc component of
the electrostatic forces, $V_{dc}$. BR excited the array near its
second instability tongue, i.e. the driving frequency lied within
the band of normal frequencies (see chapter \ref{LCmodel}).

Before applying the electrostatic forces, the characteristics of
each beam separately  were measured. The averaged fundamental
frequency was $179.3$ kHz, with a standard deviation of $0.53$ kHz,
and the quality factors $Q$ ranged from $2,000$ to $10,000$. No
correlations were found between the location of the beams in the
array and their specific properties. Therefore in the following
model (chapter~\ref{LCmodel}) the beams are regarded as identical.
Once the electrostatic forces were applied, the quality factor
values severely decreased as $V_{dc}$ was increased. This suggests
that the dissipation of the system is mainly due to induced currents
between the beams.

\begin{figure}
\begin{center}
\includegraphics[width=0.5\textwidth]{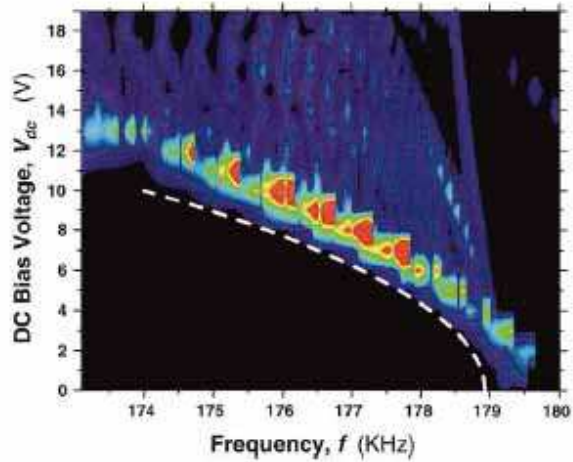}%
\caption{\label{BRresults} A color map showing relative intensities
of the diffracted light from the array, as a function of the voltage
$V_{dc}$ and the driving frequency $f$ taken from~\cite{BR02}. The
dashed white line shows a fit to the measured lower bound frequency
obtained from a simple linear theory. For $V_{dc}\approx14 V$ the
array responded at frequencies beyond the expected upper bound
frequency $f_b\approx179.3 kHz$.}
\end{center}
\end{figure}

Fig.~\ref{BRresults} shows the measured collective response of the
array as a function of the dc component of the electrostatic forces
and the driving frequency. For each value of $V_{dc}$ the amplitude
of the ac component was set to $50$ mV and the frequency was
gradually increased. The band of frequencies the array responded to
became wider as $V_{dc}$ was increased. This can be understood
already at the linear analysis level, it is a direct consequence of
the formation of the band of collective modes. The larger the dc
component is, the wider is the band. The upper bound of the band
however strongly depends on $V_{dc}$, while the linear analysis
predicts it to be simply the fundamental frequency of an individual
beam $\omega_b/2\pi\approx179.3$ kHz. Moreover, for certain values
of the dc component the array responded to frequencies beyond this
upper bound frequency. It is also evident that as the driving
frequency was swept up, the response showed a small number of wide
peaks, much wider than the 67 resonance peaks that a linear theory
predicts. These features were qualitatively explained by Lifshitz
and Cross~\cite{LC03} by taking into account the nonlinearities of
the array.

\newpage

\chapter{Model and Equations of Motion}\label{LCmodel}
Lifshitz and Cross~\cite[henceforth LC]{LC03} showed that the
nontrivial response of the array is a direct consequence of the
nonlinear character of the beams. They derived a set of equations of
motion, introducing only the essential terms for capturing the
nonlinear features obtained by BR. We shall briefly discuss some of
their main guidelines for deriving the equations of motion.

The normal frequencies of the \emph{individual} beams were found to
be well separated. Thus for moderate driving each beam is strictly
vibrating in its fundamental frequency $\omega_b$. Assuming the
beams are restricted to in-plane vibrations only, each beam can be
described by a single degree of freedom $u_n$, its displacement from
equilibrium.

For simplicity, the attractive forces due to the applied
electrostatic potential between the beams are approximated by
nearest neighbor interactions
\begin{equation}
  \label{eq:elecforce}
  F_{electric}^{(n)} = -\frac12 m\omega_b^{2}\Delta^2[1+H\cos(2\omega_p t)](u_{n+1} -
  2u_n +u_{n-1}).
\end{equation}
$\Delta^2$ and $\Delta^2H$  are the dc and ac components of the
electrostatic forces respectively, where the prefactor
$m\omega_b^2/2$ is inserted for convenience. The parametric
excitation introduced into the system by the ac component, is an
instability of the system that occurs for driving frequencies
$2\omega_p$ close to one of the special values
$2\omega_0/n$~\cite{nayfeh}, where $\omega_0$ is one of the normal
frequencies of the array and $n$ is an integer that labels the
so-called instability tongues of the system. In the BR experiment,
the system was excited in its second instability tongue, i.e.
$n=2\;,\omega_p\approx\omega_0/2$. Since the response at the second
tongue was found to be quite similar to the response at the first
tongue, the calculations preformed by LC were done for the first
instability tongue. In the current work we proceed with
concentrating on the first tongue only, and thus we always take
$\omega_p$ itself to lie within the normal frequency band.

The restoring forces of the beams are due to their elasticity.
Measurements done on individual beams indicate that the linear
restoring force is supplemented with  a cubic term of the
displacement acting to stiffen the beam. Neglecting higher order
nonlinear corrections the elastic force can therefore be written as
\begin{equation}
  \label{eq:elasticforce}
  F_{elastic}^{(n)} = -m\omega_b^{2}\left(u_n +\alpha
  u_{n}^3\right),
\end{equation}
with $\alpha>0$. Motivated by the observations of BR regarding the
quality factors (see chapter~\ref{BRexperiment}) LC assumed that the
dissipation of the system is mainly due to the electrostatic
interaction, which induces currents through the beams. Therefore the
dissipation should depend on the difference variable $u_n-u_{n-1}$,
describing the relative displacements of a pair of neighboring
beams. LC showed that a nonlinear dissipation term must also be
introduced in order to obtain bounded response for driving frequency
sweeps. Thus the dissipation terms were taken to be \footnote{The
effect of introducing gradient-dependent dissipation terms instead
of local dissipation terms is to renormalize the bare dissipation
coefficients $\Gamma$ and $\eta$ to wavenumber-dependent
coefficients of the form $4\Gamma\sin^2\left(q/2\right)$ and
$4\eta\sin^2\left(q/2\right)$ respectively, as will become apparent
below.}
\begin{eqnarray}
  \label{eq:diss}\nonumber
  &F_{diss}^{(n)}
   &= \frac12 m\omega_b\Gamma (\dot u_{n+1} - 2\dot u_n +\dot u_{n-1})\\
  &&+\ \frac12 m\omega_b\alpha\eta [(u_{n+1}-u_n)^2(\dot u_{n+1}-\dot u_n) -
            (u_{n}-u_{n-1})^2(\dot u_{n}-\dot u_{n-1})].
\end{eqnarray}
The dissipation of the system is assumed to be weak, which makes it
possible to excite the beams with relatively small driving
amplitudes. In such case the response of the beams is moderate,
justifying the description of the system with nonlinearities up to
cubic terms only. The weak dissipation can be parameterized by
introducing a small expansion parameter $\epsilon\ll1$, physically
defined by the linear dissipation coefficient
$\Gamma\equiv\epsilon\gamma$, with $\gamma$ of order one. The
driving amplitude is then expressed by $\Delta^2 H=\epsilon h$, with
$h$ of order one. The weakly nonlinear regime is studied by
expanding the displacements $u_n$ in powers of $\epsilon$. Taking
the leading term to be of the order $\epsilon^{1/2}$ ensures that
all the corrections, to a simple set of equations describing $N$
coupled harmonic oscillators, enter the equations at the same order
of $\epsilon^{3/2}$.

Introducing the scaled variables $t\rightarrow t/\omega_b$ and
$u_n\rightarrow u_n/\sqrt{\alpha}$, LC eventually obtained the
following set of dimensionless equations
\begin{eqnarray} \label{eom}\nonumber
\ddot u_n &+ &u_n + u_n^3 + \frac12\bigl[\Delta^2 + \epsilon
         h\cos(2\omega_p t)\bigr]
         (u_{n+1} - 2u_n + u_{n-1})\\ \nonumber
&- &\frac12 \epsilon\gamma
     (\dot u_{n+1} - 2\dot u_n + \dot u_{n-1})\\
&- &\frac12 \eta \bigl[(u_{n+1}-u_n)^2(\dot u_{n+1} - \dot u_n) -
(u_n-u_{n-1})^2(\dot u_n - \dot u_{n-1})\bigr] = 0.
\end{eqnarray}
The boundary conditions are set according to the experiment of BR,
who had two additional fixed beams at both ends of the array, thus
$u_0=u_{N+1}=0$.
\newpage

\chapter{Normal Modes and Linear Stability Analysis}
\label{NMandLS}

We first expand the response of the beams into $N$ standing waves,
\begin{equation} \label{StandingWaves}
u_n=\sum_{m=1}^N\Phi_m(t)\sin(q_mn),\quad{\rm with}\quad
q_m=\frac{m\pi}{N+1},\quad m=1\ldots N.
\end{equation}
Substituting (\ref{StandingWaves}) into the equations of motion
(\ref{eom}) yields $N$ nonlinear ODEs for the amplitudes of the
waves,
\begin{equation}\label{NormalModes}
\ddot{\Phi}_m+\omega_m^2\Phi_m+\epsilon 2\sin^2
(q_m/2)\left[\gamma\dot{\Phi}_m-h\cos(2\omega_p
t)\Phi_m\right]+N.L.=0,
\end{equation}
where $N.L.$ stands for nonlinear terms that couple the $N$ linear
Mathieu~\cite{LLmech} equations, and $\omega_m$ is given by the
dispersion relation
\begin{equation}\label{dispersion}
\omega^2_m=1-2\Delta^2\sin^2(q_m/2).
\end{equation}
One can easily verify that the zero-displacement state, $u_n=0$ for
all $n$, is always a solution of the equations of
motion~(\ref{eom}), though it is not always a stable one. To study
the transition to oscillating solutions, we first linearize the
equations of motion about the zero-displacement state, and introduce
a new time scale $T=\epsilon t$ upon which the growth of small
perturbations will occur. $T$ and $t$ are regarded as independent
variables, following the multiple scales procedure~\cite{nayfeh}, as
we expand the response in orders of $\epsilon$,
\begin{equation}\label{PhiExpan}
\Phi_m(t,T)=\epsilon^{1/2}\Phi_{m_0}(t,T)+
\epsilon^{3/2}\Phi_{m_1}(t,T)+O(\epsilon^{5/2}),
\end{equation}
and express the time derivatives in terms of both $t$ and $T$,
\begin{equation}\label{TDerivatives}
\partial_t\rightarrow\partial_t+\epsilon\partial_T \quad{\rm
and}\quad
\partial^2_{tt}\rightarrow\partial^2_{tt}+2\epsilon\partial^2_{tT}+O(\epsilon^2).
\end{equation}
Substituting (\ref{TDerivatives}) into the linear part of
Eq.~(\ref{NormalModes}) and collecting terms of same order
$\epsilon$ we obtain for the $\epsilon^{1/2}$ order terms
\begin{subequations}
\begin{equation}\label{Lin0order}
\left(\partial^2_{tt}+\omega_m^2\right)\Phi_{m_0}=0,
\end{equation}
and for the $\epsilon^{3/2}$ order terms
\begin{equation}\label{Lin1order}
\left(\partial^2_{tt}+\omega_m^2\right)\Phi_{m_1}=
\left(-2\partial^2_{tT}-2\gamma\sin^2\left(q_m/2\right)\partial_t+
h\sin^2\left(q_m/2\right) (e^{i2\omega_p t}+e^{-i2\omega_p
t})\right)\Phi_{m_0}.
\end{equation}
\end{subequations}
It is convenient to write the solution of Eq.~(\ref{Lin0order}) in
the following form
\begin{equation}\label{Lin0ordSol}
\Phi_{m_0}=A(T)e^{i\omega_m t}+c.c.\,,
\end{equation}
where $A(T)$ is a complex amplitude which can slowly change over
long time scales. Substituting into Eq.~(\ref{Lin1order}) we obtain
\begin{eqnarray}\label{Phi1Eq}
\left(\partial^2_{tt}+\omega_m^2\right)\Phi_{m_1}=&-&\left(
2i\omega_m\frac{d A_m}{d T}+2i\omega_m\gamma\sin^2
\left(q_m/2\right)A_m\right)e^{i\omega_m t}\\
&+&h\sin^2\left(q_m/2\right)
A_me^{i(2\omega_p+\omega_m)t}+h\sin^2\left(q_m/2\right)
A_m^*e^{i(2\omega_p-\omega_m)t}+c.c.\nonumber
\end{eqnarray}
The terms on the right hand side of Eq.~(\ref{Phi1Eq}) proportional
to $e^{i\omega_m t}$ (called \emph{secular terms}) excite the
equation for $\Phi_{m_1}$ in its resonant frequency. In order to
prevent the perturbative term $\Phi_{m_1}$ from diverging, we must
account for a \emph{solvability condition}, demanding that the sum
of all the secular terms vanishes\footnote{The constraint on the
secular terms can be obtained in a wider context of linear
differential operator theorems, as explained briefly in section
\ref{Aderivation} .}. For large frequency detunings
$\omega_p-\omega_m\gg\epsilon$, only the first two terms of the left
hand side of Eq.~(\ref{Phi1Eq}) must vanish and the solvability
condition calls
\begin{equation}\label{dampingA}
\frac{d A_m}{d T}=-\gamma\sin^2(q_m/2)A_m,
\end{equation}
yielding a slow exponential decay of $A_m$ and thus no excitation of
the $m^{th}$ mode. On the other hand for frequency detunings of
order $\epsilon$ the fourth term of the right hand side of Eq.~
(\ref{Phi1Eq}) is also secular. We introduce a detuning parameter
$\omega_p=\omega_m+\frac 12\epsilon\Omega_m$, and the solvability
condition yields
\begin{equation}\label{excitedA}
-\left(2i\omega_m\frac{d A_m}{d
T}+2i\omega_m\gamma\sin^2(q_m/2)A_m\right)+ h\sin^2(q_m/2)
A_m^*e^{i\Omega_m T}=0.
\end{equation}
Trying a solution of the form $A_m(T)=a_me^{\sigma_m
T}e^{i\frac{\Omega_m}{2}T}$, with $\sigma_m\in\mathbb{R}$ yields
\begin{eqnarray}\label{amEq}
&&-2i\omega_m\sigma_m a_m+\omega_m\Omega_m
a_m-2i\omega_m\gamma\sin^2(q_m/2)a_m+h\sin^2(q_m/2)
a_m^*=0\quad\Rightarrow\nonumber\\
&&(\omega_m\Omega_m)^2+\left(2\omega_m\gamma\sin^2(q_m/2)+2\omega_m\sigma_m\right)^2=
h^2\sin^4(q_m/2)\quad\Rightarrow\nonumber\\
&&\sigma_m=-\gamma\sin^2(q_m/2)\pm\sqrt{(h\sin^2(q_m/2)/2\omega_m)^2-(\Omega_m/2)^2}.
\end{eqnarray}
If the \emph{linear growth rate} $\sigma_m>0$ the solution will
grow, thus the critical driving amplitude $h_m(\omega_p)$ is the one
for which the growth rate $\sigma_m$ vanishes, explicitly,
\begin{equation}\label{hcm}
h_m(\omega_p)=2\omega_m\gamma\sqrt{1+\left(\frac{\Omega_m}{2\gamma\sin^2
(q_m/2)}\right)^2}=
2\omega_m\gamma\sqrt{1+\left(\frac{\omega_p-\omega_m}{\epsilon\gamma\sin^2
(q_m/2)}\right)^2}.
\end{equation}
$h_m(\omega_p)$ is the driving amplitude required to excite the
$m^{th}$ normal mode, from zero displacement into parametric
oscillations with frequency $\omega_p$. Note that it has the
familiar form of a first instability tongue~\cite{LLmech}, modified
by the dispersive correction $\sin^2(q_m/2)$.

\section{The Case of Distinct Normal Frequencies - A Single Mode
Response}
\label{SMresponse} 
\begin{figure}
\begin{center}
\includegraphics[width=0.75\textwidth]{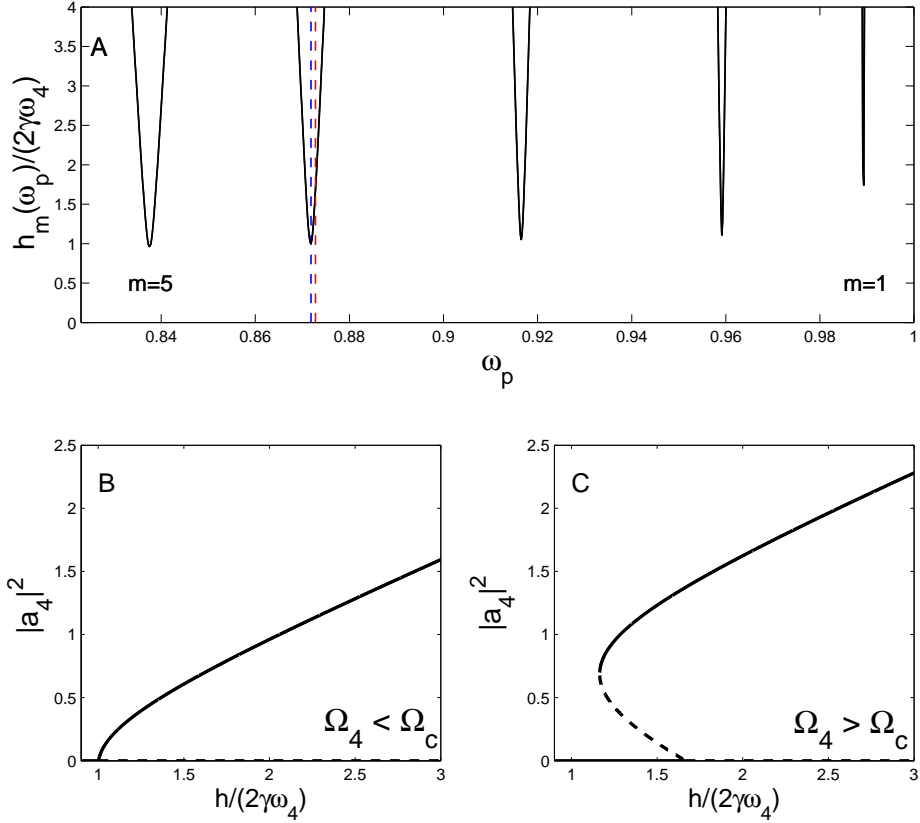}%
\caption{\label{SingleModeFig} (A) The five $h_m(\omega_p)$ curves
of an array of 5 beams in the $(h,\omega_p)$ plane are shown. Inside
the region bounded by the $m^{th}$ curve the zero displacement state
is unstable towards the growth of the $m^{th}$ mode. The dashed
vertical lines mark two excitations of the array with frequency
detunings of $\Omega_4=0.1$ (blue line) and $\Omega_4=2$ (red line).
The critical detuning is $\Omega_c=0.196$. (B) For
$\Omega_4<\Omega_c$ a supercritical bifurcation is obtained , (C)
while $\Omega_4>\Omega_c$ yields a subcritical bifurcation. The
parameters used are $\Delta=0.4,\;\eta=0.1$ and
$\epsilon\gamma=0.001$.}
\end{center}
\end{figure}

If the spacing between the normal frequencies
$\delta\omega_m=\omega_m - \omega_{m-1}$ is much greater than
$\epsilon$ one should expect the system to respond like a single
degree of freedom. Only the mode for which $\omega_p$ is close to
its normal frequency is excited. Fig. \ref{SingleModeFig}~$(A)$
shows five $h_m(\omega_p)$ curves in the $(h,\omega_p)$ plane which
mark the onset of standing waves with a wave number $q_m$ for an
array of 5 resonating beams. The curves do not overlap, indicating
that in this array for moderate drives only single modes are
excited.

In order to calculate the steady state response, a single mode
solution should be substituted into the equations of motion
(\ref{eom})
\begin{equation}\label{SMSol}
u_n=\Phi_m(t)\sin(q_mn).
\end{equation}
By taking into account the nonlinear terms neglected up to this
point, the saturation of $A_m$ can be calculated. LC performed such
a calculation and obtained the steady state of the single mode
solution
\begin{subequations}\label{SMSolution}
\begin{eqnarray}
u_n&=&\epsilon^{1/2}2|a_m|\cos(\omega_p t-\varphi)\sin(q_mn),\;{\rm
with} \nonumber\label{SMa}\\\\h^2&=&\frac{1}{\sin^4(q_m/2)}
\left(\frac94|a_m|^2-\omega_m\Omega_m\right)^2+\left(2\omega_m\gamma+
6\omega_m\eta\sin^2(q_m/2)|a_m|^2\right)^2,\;{\rm
and}\label{SMc}\nonumber\\\\\varphi&=&\frac12\arctan\left(\frac{2\sin^2(q_m/2)\gamma+
6\sin^4(q_m/2)\eta|a_m|^2}{\frac94|a_m|^2-\omega_m\Omega_m}\omega_m\right)\nonumber.\\
\label{SMb}
\end{eqnarray}
\end{subequations}
In Fig. \ref{SingleModeFig}~$(B)$ and $(C)$ the response of the
excited $m^{th}$ mode $|a_m|^2$ is plotted as a function of the
amplitude $h$ for two values of the frequency detuning $\Omega_m$.
Solid curves indicate stable solutions and dashed curves unstable
solutions. The nature of the response changes significantly as the
detuning is set above a critical value of
$\Omega_c=16\sin^6(q_m/2)\omega_m\gamma\eta/3$. For
$\Omega_m<\Omega_c$ the amplitude of the oscillating solution grows
continuously for $h$ above threshold $h>h_m(\omega_p)$ and is
stable. This is known as a supercritical Hopf
bifurcation~\cite{strogatz}. On the other hand, a subcritical Hopf
bifurcation is obtained for $\Omega_m>\Omega_c$. An unstable
solution grows below threshold, until it reaches the so-called
saddle-node point, where the curve of $|a_m|^2$ as a function of $h$
bends around (a turning point), and the solution becomes stable and
an increasing function of $h$. The stability of the steady state
solution is determined by linearizing the
solutions~(\ref{SMSolution}) about a small perturbation with the
same spatial dependence $\sin(q_m n)$ (see section~\ref{sndins}
bellow). Whether these perturbations decay or grow determines the
stability of the solution. The stability can alternatively be
deduced from a corollary of the factorization theorem \cite{ioos},
which states that the stability of the solutions switches either at
turning points or at points at which two solutions intersect. In
both approaches only instabilities towards the growth of
perturbations with the same wave number $q_m$ are examined, thus
such a stability analysis is valid only for cases of sufficiently
separated normal frequencies.

\section{The Case Overlapping Stability Curves}
\begin{figure}
\begin{center}
\includegraphics[width=0.75\textwidth]{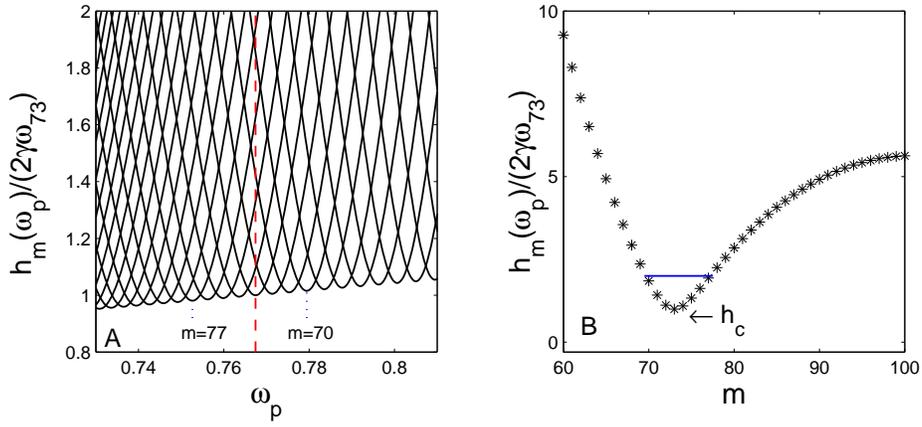}%
\caption{\label{NonDistinctFig} (A) A set of overlapping instability
tongues $h_m(\omega_p)$ for an array of $100$ oscillators as
function of $\omega_p$. Modes $64<m<81$ are plotted, the dashed
vertical line indicates the value of $\omega_p$ used in B. (B)
Neutral stability curve as a function of mode number $m$, plotted
for $\omega_p = \omega_{73}$. Marks the minimal driving amplitude
$h_m(\omega_p)$ required to excite the $m^{th}$ mode from the
zero-displacement state. The horizontal line indicates the band of
unstable modes for a particular value of $h=2$. The parameters used
are $\Delta=0.5$ and $\epsilon\gamma=0.01$.}
\end{center}
\end{figure}

If  $N$ is sufficiently large such that $\delta\omega_m \simeq
\frac{\partial\omega_m}{\partial q_m} \frac\pi{N} \sim
\frac{\Delta^2}{N} \sim \epsilon$, several modes can be excited
simultaneously for moderate driving amplitudes $h\sim O(1)$.  This
situation is illustrated for $N=100$ oscillators in
Fig.~\ref{NonDistinctFig} which shows in (A) a set of overlapping
instability tongues $h_m(\omega_p)$ for $64<m<81$ plotted as a
function of $\omega_p$, and in (B) the values of $h_m(\omega_p)$ as
a function of mode number $m$ for a particular value of
$\omega_p=\omega_{73}$ (dashed line in (A)), outlining the
\emph{neutral stability} curve below which the zero-displacement
state is stable.

In the limit $N\to\infty$ of very large arrays, the frequency
spectrum becomes essentially continuous and so does the neutral
stability curve of Fig.~\ref{NonDistinctFig} (B).  The first mode
$q_c$ to emerge when increasing the driving amplitude from zero will
then be the mode which minimizes the continuous $h_m(\omega_p)$
curve, occurring at a critical driving amplitude $h_c$. For a
moderate-size system, with a discrete frequency spectrum, the first
excited mode $q_m$ will be the one whose critical driving amplitude
$h_m(\omega_p)$ is closest to $h_c$. By further increasing the
driving amplitude, the zero-displacement state becomes unstable to a
band of wave numbers bounded by the neutral stability curve, as
indicated by the horizontal line in Fig.~\ref{NonDistinctFig} (B).
Once the excited modes start saturating into standing waves they
interact with each other through the nonlinear terms, potentially
yielding complicated dynamical behavior, as observed by
LC~\cite{LC03} in the exact solutions obtained using secular
perturbation theory. The exact solutions however could only be
obtained for small arrays, enabling only a qualitative understanding
of the dynamics of large arrays.

\newpage
\chapter{The Response of Large Arrays - The Amplitude Equation Approach}
In order to study arrays with a large number of beams, we adopt an
approach usually applied to continuous spatially-extended systems.
In such systems the emergence of modes, accessible from the unstable
band, appear as modulations of the basic pattern determined by the
initial unstable mode $q_c$. These modulations develop slowly over
long spatial scales and thus can be described by an envelope
function of the basic pattern whose dynamics obeys an appropriate
set of \emph{amplitude equations}~\cite{reviewcross}.
\section{Derivation of The Amplitude Equations}\label{Aderivation}
The general form of the amplitude equations can be deduced from
symmetry considerations, but here we wish to derive the equations
explicitly and obtain exact expressions for all the coefficients,
that can subsequently be tested quantitatively both numerically and
experimentally. We introduce a continuous displacement field
$u(x,t)$, keeping in mind that only for integral values $n$ of the
spatial coordinate does it actually correspond to the displacements
$u(x=n,t)=u_n(t)$ of the discrete set of oscillators in the array.
We also introduce slow spatial and temporal scales, $X=\epsilon x$
and $T=\epsilon t$, upon which the dynamics of the envelope function
occurs, and expand the displacement field in terms of $\epsilon$,
\begin{equation}\label{uexpansion}
u=\epsilon^{1/2}u^{(0)}(x,t,X,T)+\epsilon^{3/2}u^{(1)}(x,t,X,T)+O(\epsilon^{5/2}).
\end{equation}

In order to substitute the expansion (\ref{uexpansion}) into the
equations of motion (\ref{eom}) we must express the temporal and
spatial operators in Eq.~(\ref{eom}) in terms of the new variables
$T$ and $X$. The time derivative substitution is given by
Eq.~(\ref{TDerivatives}). The displacement of the $n\pm 1$ beam is
given by
\begin{equation}\label{ashift}
u_{n\pm1}(t)=\epsilon^{1/2}u^{(0)}(x\pm1,t,X,T)+O(\epsilon^{3/2}).
\end{equation}
We insert Eqs.~(\ref{TDerivatives}),(\ref{uexpansion}) and
(\ref{ashift}) into (\ref{eom}), and collect terms of the same order
of $\epsilon$. The $\epsilon^{1/2}$ terms yield
\begin{equation} \label{A1_2order}
\mathfrak{L} u^{(0)}=0,\qquad
\mathfrak{L}\equiv\partial_{tt}^2+\frac
12\Delta^2\left(e^{\partial_x}+e^{-\partial_x}-2\right)+1,
\end{equation}
where $e^{\partial_x}$ is an operator that performs a spatial shift
by $1$, thus the full spatial operator in Eq.~(\ref{A1_2order}) is
simply the discrete Laplacian. Eq. (\ref{A1_2order}) is solved by
setting
\begin{equation} \label{u0}
u^{(0)}(x,t,X,T)=\left(A_+(X,T)e^{-iq_px}+A_-^*(X,T)e^{iq_px}\right)e^{i\omega_p
t} +c.c.
\end{equation}
The response to the lowest order of $\epsilon$ is therefore
expressed in terms of two counter-propagating waves with modulated
complex amplitudes $A_+$ and $A_-$. This is a typical ansatz for
parametrically excited systems, though the choice of the scaling of
$X$ and $T$ with $\epsilon$ is not. The asterisk and $c.c.$ stand
for the complex conjugate and $q_p$ is determined by the driving
frequency $\omega_p$ and the dispersion relation~(\ref{dispersion})
\begin{equation}\label{dispersion_cont}
q_p=2\arcsin\sqrt{\frac{1-\omega_p^2}{2\Delta^2}}.
\end{equation}

In order to obtain the $\epsilon^{3/2}$ order contribution to the
different terms in the equations of motion, we express the
$O(\epsilon^{3/2})$ correction of Eq.~(\ref{ashift}) in terms of
$u^{(1)}$ and $A_\pm$ using the substitution
\begin{equation}\label{Aexpansion}
A_\pm(X+\epsilon,T)\simeq A_\pm(X,T)+\epsilon \frac{\partial
A_\pm(X,T)}{\partial X}.
\end{equation}
Using Eqs. (\ref{ashift}), (\ref{u0}) and (\ref{Aexpansion}) these
contributions are
\begin{subequations}
\begin{align}
&u_{n\pm1}:\nonumber\\
&\quad u^{(1)}(x\pm 1,t)\pm\left(\frac{\partial A_+}{\partial
X}e^{-iq_px}e^{\mp iq_p}+\frac{\partial A_-^*}{\partial
X}e^{iq_px}e^{\pm
iq_p}\right)e^{i\omega_p t}+c.c.\,;\\
\displaybreak[0]
&u_{n+1}+u_{n-1}-2u_n:\nonumber\\
&\quad u^{(1)}(x+1,t)+u^{(1)}(x-1,t)-2u^{(1)}(x,t)\nonumber\\&\quad+
(e^{-iq_p}-e^{iq_p})\left(\frac{\partial A_+}{\partial
X}e^{-iq_px}-\frac{\partial A_-^*}{\partial
X}e^{iq_px}\right)e^{i\omega_p t}+c.c.\nonumber\\&=
u^{(1)}(x+1,t)+u^{(1)}(x-1,t)-2u^{(1)}(x,t)\nonumber\\&\quad-
2i\sin(q_p)\left(\frac{\partial A_+}{\partial
X}e^{-iq_px}-\frac{\partial A_-^*}{\partial
X}e^{iq_px}\right)e^{i\omega_p
t}+c.c.\,;\\
\displaybreak[0]
&\epsilon(\dot u_{n+1}+\dot u_{n-1}-2\dot
u_n):\nonumber\\&\quad-4i\omega_p\sin^2(q_p/2)\left(A_+e^{-iq_px}+A_-^*e^{iq_px}\right)e^{i\omega_p
t}+c.c.\,;\\ \nonumber\\ \displaybreak[0]
&\epsilon\cos(2\omega_p
t)(u_{n+1}+u_{n-1}-2u_n):\nonumber\\&\quad-2\sin^2(q_p/2)\left(A_+e^{-iq_px}+A_-^*e^{iq_px}
\right)e^{i(2\omega_p+\omega_p)t}\nonumber\\&\quad
-2\sin^2(q_p/2)\left(A_+^*e^{iq_px}+A_-e^{-iq_px}
\right)e^{i(2\omega_p-\omega_p)t}+c.c.\,;\\
\displaybreak[0]
&u_n^3:\nonumber\\&\quad\;3|A_+e^{-iq_px}+A_-^*e^{iq_px}|^2\left(A_+e^{-iq_px}+A_-^*e^{iq_px}
\right)e^{i\omega_p t}+O\left(e^{i3\omega_p
t},e^{i3q_px}\right)+c.c.\nonumber\\&=
3(|A_+|^2+2|A_-|^2)A_+e^{-iq_px}e^{i\omega_p t}+
3(2|A_+|^2+|A_-|^2)A_-^*e^{iq_px}e^{i\omega_p
t}\nonumber\\&\quad+O\left(e^{i3\omega_p
t},e^{i3q_px}\right)+c.c.\,;\\
\displaybreak[0]
&(u_{n\pm1}-u_n)^2(\dot u_{n\pm1}-\dot u_n)=\frac13\frac d{dt}(u_{n\pm1}-u_n)^3:\nonumber\\
&\quad\;\frac13\frac d{dt}\left((A_+e^{-iq_px}(e^{\mp
iq_p}-1)+A_-^*e^{iq_px}(e^{\pm iq_p}-1))e^{i\omega_p
t}+c.c.\right)^3\nonumber\\&=\frac d{dt}|A_+e^{-iq_px}(e^{\mp
iq_p}-1)+A_-^*e^{iq_px}(e^{\pm
iq_p}-1)|^2\nonumber\\&\quad\;\times\left(A_+e^{-iq_px}(e^{\mp
iq_p}-1)+A_-^*e^{iq_px}(e^{\pm iq_p}-1)\right)e^{i\omega_p
t}+O(e^{i3\omega_p t},e^{i3q_px})+c.c.\nonumber\\
&=i\omega_p\left[\vphantom{A_-^*e^{i2q_px}}\right.|e^{-iq_p}-1|^2(|A_+|^2+|A_-|^2)+A_+A_-e^{-i2q_px}(e^{\mp
iq_p}-1)^2\nonumber\\&\quad\;+A_+^*A_-^*e^{i2q_px}(e^{\pm
iq_p}-1)^2\left.\vphantom{A_-^*e^{i2q_px}}\right]\left(A_+e^{-iq_px}(e^{\mp
iq_p}-1)+A_-^*e^{iq_px}(e^{\pm iq_p}-1)\right)e^{i\omega_p
t}\nonumber\\
&\quad\; +O(e^{i3\omega_p t},e^{i3q_px})+c.c.\,;\\
\nonumber\displaybreak[0]
{\rm and}\\
&(u_{n+1}-u_n)^2(\dot u_{n+1}-\dot u_n)+ (u_{n-1}-u_n)^2(\dot
u_{n-1}-\dot u_n):\nonumber\\&\quad\;
i\omega_p|e^{-iq_p}-1|^2(e^{iq_p}+e^{-iq_p}-2)(|A_+|^2+2|A_-|^2)A_+
e^{-iq_px}e^{i\omega_p t}\nonumber\\&\quad\;
+i\omega_p|e^{-iq_p}-1|^2(e^{iq_p}+e^{-iq_p}-2)(2|A_+|^2+|A_-|^2)A_-^*
e^{iq_px}e^{i\omega_p t}\nonumber\\&\quad\;+O(e^{i3\omega_p
t},e^{i3q_px})+c.c.\nonumber\\&=-i16\omega_p
\sin^4(q_p/2)\left[\vphantom{A_-^*e^{i2q_px}}\right.(|A_+|^2+2|A_-|^2)A_+
e^{-iq_px}e^{i\omega_p
t}+\\&\quad\;\left.(2|A_+|^2+\nonumber|A_-|^2)A_-^*
e^{iq_px}e^{i\omega_p t}\right]+O(e^{i3\omega_p
t},e^{i3q_px})+c.c.\,,
\end{align}
\end{subequations}
where $O(e^{i3\omega_p t},e^{i3q_px})$ are terms proportional to
$e^{i3\omega_p t}$ or $e^{i3q_px}$ which do not enter the dynamics
at the lowest order of the $\epsilon$ expansion. The
$\epsilon^{3/2}$ order terms of the equations of motion~(\ref{eom})
therefore yield
\begin{eqnarray}\label{A3_2order}
\mathfrak{L}u^{(1)}&=&f\nonumber\\
&=&
\left[-2i\omega_p\left(\frac{\partial A_+}{\partial T}e^{-iq_px}+
\frac{\partial A_-^*}{\partial
T}e^{iq_px}\right)+i\sin(q_p)\Delta^2\left(\frac{\partial
A_+}{\partial X}e^{-iq_px}- \frac{\partial A_-^*}{\partial
X}e^{iq_px}\right)\right]e^{i\omega_p t}\nonumber\\&&+\left[h
\left(A_+^*e^{iq_px}+A_-e^{-iq_px}\right)-i2\omega_p\gamma
\left(A_+e^{-iq_px}+A_-^*e^{iq_px}\right)\right]\sin^2(q_p/2)e^{i\omega_p
t}\nonumber\\&&-
\left(3+i8\eta\omega_p\sin^4(q_p/2)\right)\left(|A_+|^2+2|A_-|^2\right)
A_+e^{-iq_px}e^{i\omega_p t}\nonumber\\&&-
\left(3+i8\eta\omega_p\sin^4(q_p/2)\right)\left(2|A_+|^2+|A_-|^2\right)
A_-^*e^{iq_px}e^{i\omega_p t} + O(e^{i3\omega_p
t},e^{i3q_px})+c.c.\nonumber\\&&
\end{eqnarray}
Let us denote by $V_0$ the zero eigenvectors of the self adjoint
operator $\mathfrak{L}$, namely, $e^{-i(q_px-\omega_p t)}$ and
$e^{i(q_px+\omega_p t)}$ and their complex conjugates. If $u^{(1)}$
is a solution of Eq.~(\ref{A3_2order}), then $f$ must be orthogonal
to the zero eigenvectors because
\begin{equation}\label{Fredholm}
(f,V_0)=(\mathfrak{L}u^{(1)},V_0)=(u^{(1)},\mathfrak{L}V_0)=0,
\end{equation}
where $(f,g)\propto\int f(x,t)g^*(x,t)dxdt$ is the inner product of
the $L_2(\mathbb{C})$ space.  The Fredholm alternative
theorem~\cite{stakgold} states that Eq.~(\ref{Fredholm}) is also a
sufficient condition for the existence of a solution $u^{(1)}$ for
Eq.~(\ref{A3_2order}). Thus the solvability condition of
Eq.~(\ref{A3_2order}) requires that the coefficients of the zero
eigenvector terms at its right hand side must vanish identically. We
therefore obtain two coupled amplitude equations,
\begin{subequations}\label{AmpEqs}
\begin{align}
\frac{\partial A_+}{\partial T}+v_g\frac{\partial A_{+}}{\partial
X}&= -\gamma\sin^{2}(q_p/2)A_{+}-i\frac{h}
{2\omega_p}\sin^2(q_p/2)A_{-}\\&
-\left(4\eta\sin^4(q_p/2)-i\frac{3}{2\omega_p}\right)
\left(|A_{+}|^{2}+2|A_{-}|^{2}\right)A_{+},\nonumber
\\\frac{\partial A_-}{\partial T}-v_g\frac{\partial
A_{-}}{\partial X}&=-\gamma\sin^2(q_p/2)A_{-}+ i\frac{h} {2\omega_p}
\sin^2(q_p/2)A_{+}\\&
-\left(4\eta\sin^4(q_p/2)+i\frac{3}{2\omega_p}\right)\left(2|A_{+}|^{2}+|A_{-}|^{2}\right)A_{-},
\nonumber
\end{align}
\end{subequations}
where $v_g = \frac{\partial\omega}{\partial q} =
-\frac{\Delta^{2}\sin(q_p)}{2\omega_p}$ is the usual group velocity.
Eqs.~(\ref{AmpEqs}) are two coupled complex Ginzburg-Landau
equations (CGLE)~\cite{AL02}, similar to the amplitude equations
previously derived for describing Faraday waves
excitations~\cite{ezerskii,milner}.

\section{Linear Stability Analysis of The Zero-Displacement
State}\label{LinearStabiltyAeq}
We now wish to reexamine the stability of the zero-displacement
state, using the amplitude equations (\ref{AmpEqs}) we derived. For
small perturbations of the zero displacement state
\begin{equation}
A_\pm=\alpha_\pm(T) e^{-ikX},\qquad |\alpha_\pm|\ll 1,
\end{equation}
only the linear terms of Eqs.~(\ref{AmpEqs}) are considered. Thus we
can write the amplitude equations in a matrix form
\begin{equation}\label{linearAeq}
\frac{\partial}{\partial T}
\left(%
\begin{array}{c}
  \alpha_+ \\
  \alpha_- \\
\end{array}%
\right) =
\left(%
\begin{array}{cc}
  iv_gk-\gamma\sin^2(q_p/2) &
  -i\frac{h}{2\omega_p}\sin^2(q_p/2) \\
  i\frac{h}{2\omega_p}\sin^2(q_p/2) &
  -iv_gk-\gamma\sin^2(q_p/2)\\
\end{array}%
\right)
\left(%
\begin{array}{c}
  \alpha_+ \\
  \alpha_- \\
\end{array}%
\right),
\end{equation}
describing the dynamics at the onset of modes with a wave number
$q=q_p+\epsilon k$. The solutions of Eq.~(\ref{linearAeq}) can be
expressed as linear combinations of two exponents in time,
$Be^{\sigma_+ T}$ and $De^{\sigma_- T}$, with growth rates
$\sigma_\pm$ determined by the eigenvalues of the matrix in the
right hand side of the equation. These are given by
\begin{align}\label{growthrates}
&\sigma_\pm(h)=-\gamma\sin^2(q_p/2)\pm\sqrt{(h\sin^2(q_p/2)/2\omega_p)^2-(v_gk)^2}.
\end{align}
The critical amplitude $h_k(\omega_p)$ required to excite
oscillations with a wave number $q_p+\epsilon k$ is obtained by
setting the larger eigenvector $\sigma_+(h_k)=0$, thus
\begin{equation}\label{hk}
h_k(\omega_p)=2\gamma\omega_p\sqrt{1+\left(\frac{v_gk}{\gamma\sin^2(q_p/2)}\right)^2}.
\end{equation}
The minimum of $h_k(\omega_p)$, $h_c=2\gamma\omega_p$, is obtained
for $k=0$, i.e. the amplitude equations approach yields an initial
instability with a wave number $q_c=q_p$. Comparing Eq.~(\ref{hk})
with the neutral stability curve Eq.~(\ref{hcm}) obtained by a
direct linear stability analysis of the equations of
motion~(\ref{eom}) reveals the scope of accuracy of the scalings we
use.



For $\frac 1N\ll\epsilon$, the spectrum of the normal modes can be
regarded as being essentially continuous, by replacing $q_m$ and
$\omega_m$ by continuous variables $q$ and $\omega$ related by a
continuous dispersion relation~(\ref{dispersion}).
In this limit, with $\omega_p$ taken from the normal frequency band,
$q_c$ is determined by the minimum of the continuous neutral
stability curve Eq.~(\ref{hcm}),
\begin{equation}\label{hcm_cont}
h(\omega,\omega_p)=2\gamma\omega\sqrt{1+\left(\frac{\omega-\omega_p}
{\epsilon\gamma\sin^2(q/2)}\right)^2}.
\end{equation}
Even though the beams oscillate at frequency $\omega_p$, the initial
unstable wave number $q_c$ is only approximately $q_p$. We focus on
small detunings $\Omega=2(\omega_p-\omega)/\epsilon\ll1$ and expand
Eq.~(\ref{hcm_cont}) in $\Omega$,
\begin{eqnarray}\label{hcm_expand}
h(\Omega,\omega_p)&=&2\gamma(\omega_p-\epsilon\Omega/2)
\sqrt{1+\left(\frac{\Omega}{2\gamma\sin^2(q/2)}\right)^2}
\nonumber\\&=&2\gamma(\omega_p-\epsilon\Omega/2)
\left(1+\frac12\left(\frac{\Omega}{2\gamma\sin^2(q/2)}\right)^2\right)+O(\Omega^4)
\\&=&2\gamma\omega_p\left(1-\frac{\epsilon\Omega}{2\omega_p}+\frac12
\left(\frac{\Omega}{2\gamma\sin^2(q/2)}\right)^2\right)+O(\Omega^4,\epsilon\Omega^3)
\nonumber\\&=&2\gamma\omega_p\left(1+\frac12\left(\frac{\Omega}{2\gamma\sin^2(q/2)}-
\frac{\epsilon\gamma\sin^2(q/2)}{\omega_p}\right)^2\right)-
\frac{\epsilon^2\gamma^3\sin^4(q/2)}{2\omega_p}+O(\Omega^4,\epsilon\Omega^3).
\nonumber
\end{eqnarray}
Identifying the frequency detuning with
\begin{equation}
\epsilon\Omega/2=\omega_p-\omega=|v_g|(q-q_p)=\epsilon |v_g|k,
\end{equation}
and comparing Eq.~(\ref{hcm_expand}) with the expansion of
Eq.~(\ref{hk}) in small $k$ reveals that in the amplitude equation
approach the wave number detuning $\epsilon k$ corresponds to
$q-q_p$ up to $\epsilon^2$ order corrections. This suggests that the
scaling we chose for the spatial coordinate $X=\epsilon x$ is valid
for the emergence of waves with $q-q_p\gg\epsilon^2$.

For $h=h_k(\omega_p)$ the two eigenvalues are $\sigma_+=0$ and
$\sigma_-=-2\gamma\sin^2{q_p/2}$, with the corresponding
eigenvectors being
\begin{equation}\label{EigenVectors}
V_+=
\left(%
\begin{array}{c}
  1 \\
  e^{i\psi} \\
\end{array}%
\right)\quad{\rm and}\quad V_-=
\left(%
\begin{array}{c}
  1 \\
  e^{-i\psi} \\
\end{array}%
\right),\quad{\rm where}\quad e^{i\psi}\equiv
i\frac{\gamma\sin^2(q_p/2)-iv_gk}{\sqrt{\gamma^2\sin^4(q_p/2)+(v_gk)^2}}.
\end{equation}
The amplitudes $B$ and $D$ can be expressed in terms of $\alpha_\pm$
by multiplying Eq.~(\ref{linearAeq}) by the matrix
\begin{equation}\label{EigenMatrix}
\left(%
\begin{array}{cc}
  1 &
  1 \\
  e^{i\psi} &
  e^{-i\psi} \\
\end{array}%
\right)^{-1}=\frac i{2\sin(\psi)}
\left(%
\begin{array}{cc}
  e^{-i\psi} &
  -1 \\
  -e^{i\psi} &
  1 \\
\end{array}%
\right)
\end{equation}
from the left. We then obtain that $B$ and $D$ are given by
\begin{equation}
\left(%
\begin{array}{c}
  B \\
  D \\
\end{array}
\right)\equiv
\left(%
\begin{array}{c}
  e^{-i\psi}\alpha_+-\alpha_- \\
  -e^{i\psi}\alpha_++\alpha_- \\
\end{array}
\right),
\end{equation}
and obey
\begin{equation}
\frac\partial{\partial T}
\left(%
\begin{array}{c}
  B \\
  D \\
\end{array}
\right)=
-2\gamma\sin^2(q_p/2)\left(%
\begin{array}{c}
  0 \\
  D \\
\end{array}
\right),
\end{equation}
as expected. At the onset of oscillations with a wave number
$q_p+\epsilon k$ the linear combination $B$ becomes unstable while
$D$ decays exponentially with a growth rate
$\sigma_-=-2\gamma\sin^2(q_p/2)$, suggesting that the dynamics close
to the onset of oscillations can be well described using a single
complex amplitude. This is what we do in the next section.

\newpage
\chapter{Reduction to a Single Amplitude Equation}
Motivated by the linear analysis of Eqs.~(\ref{AmpEqs}) preformed in
chapter~\ref{LinearStabiltyAeq}, we express the amplitudes $A_\pm$
in terms of $B$ and $D$,
\begin{equation}
\left(%
\begin{array}{c}
  A_+ \\
  A_- \\
\end{array}
\right)=
\left(%
\begin{array}{cc}
  1 &
  1 \\
  e^{i\psi} &
  e^{-i\psi} \\
\end{array}%
\right)
\left(%
\begin{array}{c}
  B \\
  D \\
\end{array}
\right).
\end{equation}
Just above threshold $h-h_c\ll h_c$ the $D$ amplitude decays
exponentially in time. After transients that last over periods of
order $\Gamma^{-1}\sim\epsilon^{-1}$, the response can therefore be
expressed by the $B$ amplitude only
\begin{equation}\label{A2B}
\left(%
\begin{array}{c}
  A_+ \\
  A_- \\
\end{array}
\right)=
\left(%
\begin{array}{c}
  1 \\
  e^{i\psi} \\
\end{array}
\right)B.
\end{equation}
$B(X,T)=|B(X,T)|e^{i\varphi_B(X,T)}$ represents an envelope of a
single standing wave as apparent from substituting (\ref{A2B}) back
into (\ref{u0})
\begin{eqnarray}\label{u0withB}
u^{(0)}(x,t,X,T)&=&\left(B(X,T)e^{-iq_px}+B(X,T)^*e^{i(q_px-\psi)}\right) e^{i\omega_pt}+c.c.\nonumber\\
&=&2|B(X,T)|\left(\cos\left(\omega_p t-q_px+\varphi_B\right)+
\cos\left(\omega_p t+q_px-\varphi_B-\psi\right)\right)\nonumber\\
&=&4|B(X,T)|\cos(\omega_p t-\psi/2)\cos(q_px-\varphi_B(X,T)-\psi/2).
\end{eqnarray}
$\psi/2$ is the temporal phase relative to the drive.

This reduction of the description of the dynamics to a single
amplitude $B$ is similar to the procedure introduced by
Riecke~\cite{riecke} for describing the onset of Faraday waves. It
is typical of parametric driving, which excites a single standing
wave at threshold.
\section{Scaling of $A_\pm$ Just Above Threshold}
We define a reduced driving amplitude $g$ by letting
\begin{equation}\label{Def_g}
(h-h_c)/h_c\equiv g\delta,\qquad\delta\ll 1.
\end{equation}
In order to obtain an equation, describing the relevant slow dynamics of the new
amplitude $B$, we need to select the proper scaling of the original
amplitudes $A_\pm$, as well as their spatial and temporal variables,
with respect to the new small parameter $\delta$.

If the coefficient of nonlinear dissipation $\eta$ is of order of
one, it is apparent from the original amplitude
equations~(\ref{AmpEqs}) that the cubic terms saturate the growth of
the amplitudes $A_\pm$. However, it is physically realistic to
assume that $\eta$ is small. Therefore a quintic term must enter in
order to saturate the growth of the amplitudes $A_\pm$.  This can be
achieved by defining the small parameter $\delta$ with respect to
the coefficient of nonlinear dissipation, letting
\begin{equation}\label{eta_0}
\eta=\delta^{1/2}\eta_0,\qquad\eta_0\sim O(1),
\end{equation}
and taking the amplitudes to be of order $\delta^{1/4}$.

It is apparent from Eq.~(\ref{growthrates}) for the linear growth
rates that with a drive amplitude that scales like $\delta$ the
growth rate of the amplitude $B$ scales like $\delta$ as well. The
bandwidth of unstable wave numbers scales as $\delta^{1/2}$, as
obtained by Eq.~(\ref{hk}). The new temporal and spatial scales are
therefore defined by $\tau=\delta T$ and $\xi=\delta^{1/2}X$
respectively, and we finally make the ansatz that
\begin{equation}\label{Bansatz}
\left(%
\begin{array}{c}
  A_+ \\
  A_- \\
\end{array}%
\right)=\delta^{1/4}
\left(%
\begin{array}{c}
  1 \\
  i \\
\end{array}%
\right) B(\xi,\tau)+
\delta^{3/4}\left(%
\begin{array}{c}
  w^{(1)}(X,T,\xi,\tau) \\
  v^{(1)}(X,T,\xi,\tau) \\
\end{array}%
\right)+
\delta^{5/4}\left(%
\begin{array}{c}
  w^{(2)}(X,T,\xi,\tau) \\
  v^{(2)}(X,T,\xi,\tau) \\
\end{array}%
\right).
\end{equation}
The phase $e^{i\psi}$ defined by~(\ref{EigenVectors}) to the lowest
order of $\delta$ is $i$, because the band of unstable wave numbers
$k$ scales like $\delta^{1/2}$.

\section{Derivation of The B Amplitude Equation}
The amplitude equation for $B(\xi,\tau)$ is derived by once again
using the multiple scales method. We substitute Eqs.~(\ref{Def_g}),
(\ref{eta_0}) and the ansatz~(\ref{Bansatz}) into the original
amplitude equations (\ref{AmpEqs}), and collect terms of the same
order of $\delta$. The $\delta^{1/4}$ order terms satisfy
Eqs.~(\ref{AmpEqs}) trivially, due to the fact that the
$\delta^{1/4}$ order terms in the ansatz~(\ref{Bansatz}) were chosen
according to the linear stability analysis of Eqs.~(\ref{AmpEqs})
performed in section~\ref{LinearStabiltyAeq}.

In order to obtain the higher order terms in $\delta$ we perform the
following calculations
\begin{equation}\label{deltacorrections}
\begin{split}
&|A_{+}|^{2}=\delta^{1/2}|B|^{2}+\delta(B{w^{(1)}}^*+B^*w^{(1)})\,;\\
&|A_{-}|^{2}=\delta^{1/2}|B|^{2}+
\delta(iBv{^{(1)}}^*-iB^*v^{(1)})\,;\\&
|A_{+}|^{2}A_{+}=\delta^{3/4}|B|^{2}B+\delta^{5/4}(B^{2}{w^{(1)}}^*+2|B|^{2}w^{(1)})\,;\\&
|A_{-}|^{2}A_{+}=\delta^{3/4}|B|^{2}B+\delta^{5/4}(iB^{2}{v^{(1)}}^*-i|B|^{2}v^{(1)}+|B|^{2}w^{(1)})\,;\\&
|A_{+}|^{2}A_{-}=\delta^{3/4}|B|^{2}iB+\delta^{5/4}(iB^{2}{w^{(1)}}^*+i|B|^{2}w^{(1)}+|B|^{2}v^{(1)})\,;\\&
|A_{-}|^{2}A_{-}=\delta^{3/4}|B|^{2}iB+\delta^{5/4}(-B^{2}{v^{(1)}}^*+2|B|^{2}v^{(1)})\,;\\&
\frac{\partial A_{+}}{\partial T}=\delta^{3/4}\frac{\partial
w^{(1)}}{\partial T}+\delta^{5/4}\frac{\partial w^{(2)}}{\partial
T}+\delta^{5/4}\frac{\partial B}{\partial \tau}\,;\\&\frac{\partial
A_{-}}{\partial T}=\delta^{3/4}\frac{\partial v^{(1)}}{\partial
T}+\delta^{5/4}\frac{\partial v^{(2)}}{\partial T}+\delta^{5/4}i
\frac{\partial B}{\partial \tau}\,;\\& \frac{\partial
A_{+}}{\partial X}=\delta^{3/4}\frac{\partial w^{(1)}}{\partial
X}+\delta^{3/4}\frac{\partial B}{\partial
\xi}+\delta^{5/4}\frac{\partial w^{(2)}}{\partial
X}+\delta^{5/4}\frac{\partial
w^{(1)}}{\partial\xi}\,;\\&\frac{\partial A_{-}}{\partial
X}=\delta^{3/4}\frac{\partial v^{(1)}}{\partial
X}+\delta^{3/4}i\frac{\partial B}{\partial
\xi}+\delta^{5/4}\frac{\partial v^{(2)}}{\partial
X}+\delta^{5/4}\frac{\partial v^{(1)}}{\partial\xi}\,.
\end{split}
\end{equation}
Collecting all the $\delta^{3/4}$ order terms of Eqs.~(\ref{AmpEqs})
yields
\begin{equation}\label{B3_4order}
\mathfrak{O}\left(%
\begin{array}{c}
  w^{(1)} \\
  v^{(1)} \\
\end{array}%
\right)=\left(-v_g\frac{\partial B}{\partial
\xi}+i\frac9{2\omega_p}|B|^2B\right)
\left(%
\begin{array}{c}
  1 \\
  -i \\
\end{array}%
\right),
\end{equation}
where $\mathfrak{O}$ is a linear operator given by
\begin{equation} \label{B_Operator}
\mathfrak{O}\equiv\left(%
\begin{array}{cc}
  \partial_T+v_g\partial_X+\gamma\sin^2(q_p/2) & i\gamma\sin^2(q_p/2) \\
  -i\gamma\sin^2(q_p/2) & \partial_T-v_g\partial_X+\gamma\sin^2(q_p/2) \\
\end{array}%
\right).
\end{equation}
There is no solvability condition since the right hand side of
Eq.~(\ref{B3_4order}) is automatically orthogonal to the zero
eigenvalue of $\mathfrak{O}$, $\scriptsize\left(%
\begin{array}{c}
  1 \\
  i \\
\end{array}%
\right)$. The solution of Eq.~(\ref{B3_4order}) is given by
\begin{equation}\label{w1v1Sol}
\left(\begin{array}{c}
  w^{(1)} \\
  v^{(1)} \\
\end{array}%
\right)=\frac1{2\gamma\sin^2(q_p/2)}\left(-v_g\frac{\partial B}{\partial
\xi}+i\frac9{2\omega_p}|B|^2B\right) \left(
\begin{array}{c}
  1 \\
  -i \\
\end{array}%
\right).
\end{equation}
We plug Eq.~(\ref{w1v1Sol}) back into Eqs.~(\ref{AmpEqs}) using
Eqs.~(\ref{deltacorrections}), collect all the terms with
$\delta^{5/4}$  and obtain
\newpage
\begin{subequations}
\begin{eqnarray}
&&\left[\partial_T+v_g\partial_X+\gamma\sin^2(q_p/2)\right]w^{(2)}
+i\gamma\sin^2(q_p/2)v^{(2)}\nonumber\\
&&=-\frac{\partial B}{\partial\tau}-v_g\frac{\partial
w^{(1)}}{\partial\xi}+\gamma\sin^2(q_p/2)gB-12\eta_0\sin^4(q_p/2)|B|^2B\nonumber\\
&&\quad+i\frac3{2\omega_p}\left(B^2{w^{(1)}}^*+2|B|^2w^{(1)}+
2iB^2{v^{(1)}}^*-2i|B|^2v^{(1)}+2|B|^2w^{(1)}\right)\nonumber\\
&&=-\frac{\partial
B}{\partial\tau}-\frac{v_g}{\gamma\sin^2(q_p/2)}\frac{\partial}{\partial\xi}
\left(\frac{-v_g}{2}\frac{\partial
B}{\partial\xi}+i\frac9{4\omega_p}|B|^2B\right)
+\gamma\sin^2(q_p/2)gB-12\eta_0\sin^4(q_p/2)|B|^2B\nonumber\\
&&\quad+i\frac3{2\gamma\sin^2(q_p/2)\omega_p}\left[B^2\left(\frac{-v_g}{2}\frac{\partial
B^*}{\partial\xi}-i\frac9{4\omega_p}|B|^2B^*\right)+4|B|^2
\left(\frac{-v_g}{2}\frac{\partial
B}{\partial\xi}+i\frac9{4\omega_p}|B|^2B\right)\vphantom{1}\right.\nonumber\\
&&\quad+2iB^2\left(-i\frac{v_g}{2}\frac{\partial B^*}{\partial
\xi}+\frac9{4\omega_p}|B|^2B^*\right)-2i|B|^2\left(i\frac{v_g}{2}\frac{\partial
B}{\partial\xi}+\frac9{4\omega_p}|B|^2B\right)\left.\vphantom{\frac12}\right]\nonumber\\
&&=-\frac{\partial
B}{\partial\tau}+\frac{v_g^2}{2\gamma\sin^2(q_p/2)}\frac{\partial^2
B}{\partial\xi^2}-i\frac{9v_g}{4\omega_p\gamma\sin^2(q_p/2)}\left(2|B|^2\frac{\partial
B}{\partial\xi}+B^2\frac{\partial B^*}{\partial\xi}\right)
+\gamma\sin^2(q_p/2)gB\nonumber\\&&\quad-12\eta_0\sin^4(q_p/2)|B|^2B
+i\frac{v_g}{\omega_p\gamma\sin^2(q_p/2)}\left(-3+\frac32\right)|B|^2\frac{\partial
B}{\partial\xi}\nonumber\\&&\quad
+i\frac{v_g}{\omega_p\gamma\sin^2(q_p/2)}\left(-\frac34+\frac32\right)B^2\frac{\partial
B^*}{\partial\xi}
+\frac1{\omega_p^2\gamma\sin^2(q_p/2)}\left(\frac{27}{8}-\frac{27}2-\frac{27}4+\frac{27}4\right)|B|^4B\nonumber\\
&&=-\frac{\partial
B}{\partial\tau}+\frac{v_g^2}{2\gamma\sin^2(q_p/2)}\frac{\partial^2
B}{\partial\xi^2}+\gamma\sin^2(q_p/2)gB-12\eta_0\sin^4(q_p/2)|B|^2B\nonumber\\
&&\quad-i\frac{3v_g}{2\omega_p\gamma\sin^2(q_p/2)}\left(4|B|^2\frac{\partial
B}{\partial\xi}+B^2\frac{\partial
B^*}{\partial\xi}\right)-\frac{81}{8\omega_p^2\gamma\sin^2(q_p/2)}|B|^4B,
\end{eqnarray}
and
\begin{eqnarray}
&&\left[\partial_T-v_g\partial_X+\gamma\sin^2(q_p/2)\right]v^{(2)}
-i\gamma\sin^2(q_p/2)w^{(2)}\nonumber\\
&&=-i\frac{\partial B}{\partial\tau}+v_g\frac{\partial
v^{(1)}}{\partial\xi}+\gamma\sin^2(q_p/2)giB-12\eta_0\sin^4(q_p/2)|B|^2iB\nonumber\\
&&\quad-i\frac3{2\omega_p}\left(2iB^2{w^{(1)}}^*+2i|B|^2w^{(1)}
+2|B|^2v^{(1)}-B^2{v^{(1)}}^*+2|B|^2v^{(1)}\right)\nonumber\\
&&=-i\frac{\partial B}{\partial\tau}-iv_g\frac{\partial
w^{(1)}}{\partial\xi}+\gamma\sin^2(q_p/2)giB-12\eta_0\sin^4(q_p/2)|B|^2iB\nonumber\\
&&\quad-i\frac3{2\omega_p}\left(2B^2{v^{(1)}}^*-2|B|^2v^{(1)}
-4|B|^2iw^{(1)}-iB^2{w^{(1)}}^*\right)\nonumber\\
&&=i\left[-\frac{\partial B}{\partial\tau}-v_g\frac{\partial
w^{(1)}}{\partial\xi}+\gamma\sin^2(q_p/2)gB-12\eta_0\sin^4(q_p/2)|B|^2B\vphantom{1}\right.\nonumber\\
&&\quad-i\frac3{2\omega_p}\left(-2iB^2{v^{(1)}}^*+2i|B|^2v^{(1)}
-4|B|^2w^{(1)}-B^2{w^{(1)}}^*\right)\left.\vphantom{\frac12}\right]\nonumber\\
&&=i\left[-\frac{\partial
B}{\partial\tau}+\frac{v_g^2}{2\gamma\sin^2(q_p/2)}\frac{\partial^2
B}{\partial\xi^2}+\gamma\sin^2(q_p/2)gB-12\eta_0\sin^4(q_p/2)|B|^2B\vphantom{1}\right.\nonumber\\
&&\quad-i\frac{3v_g}{2\omega_p\gamma\sin^2(q_p/2)}\left(4|B|^2\frac{\partial
B}{\partial\xi}+B^2\frac{\partial
B^*}{\partial\xi}\right)-\frac{81}{8\omega_p^2\gamma\sin^2(q_p/2)}|B|^4B\left.\vphantom{\frac12}\right].
\end{eqnarray}
\end{subequations}
In matrix form, these equations become
\begin{eqnarray}\label{B5_4order}
&&\mathfrak{O}\left(%
\begin{array}{c}
  w^{(2)} \\
  v^{(2)} \\
\end{array}%
\right)=\left[-\frac{\partial
B}{\partial\tau}+\frac{v_g^2}{2\gamma\sin^2(q_p/2)}\frac{\partial^2
B}{\partial\xi^2}+\gamma\sin^2(q_p/2)gB-12\eta_0\sin^4(q_p/2)|B|^2B\vphantom{1}\right.\nonumber\\
&&+i\frac{3(-v_g)}{2\omega_p\gamma\sin^2(q_p/2)}\left(4|B|^2\frac{\partial
B}{\partial\xi}+B^2\frac{\partial
B^*}{\partial\xi}\right)-\frac{81}{8\omega_p^2\gamma\sin^2(q_p/2)}|B|^4B\left.\vphantom{\frac12}\right]
\left(%
\begin{array}{c}
  1 \\
  i \\
\end{array}%
\right).
\end{eqnarray}
The solvability condition of Eq.~(\ref{B5_4order}) determines the
amplitude equation for $B$. Again, the Fredholm alternative theorem
states that the zero
eigenvalue of the operator $\mathfrak{O}$, $\scriptsize\left(%
\begin{array}{c}
  1 \\
  i \\
\end{array}%
\right)$, must be orthogonal to the right hand side of the equation.
Therefore, $B$ must satisfy
\begin{multline}\label{BampEg}
\frac{\partial
B}{\partial\tau}=\gamma\sin^2(q_p/2)gB+\frac{v_g^2}{2\gamma\sin^2(q_p/2)}\frac{\partial^2
B}{\partial\xi^2}-12\eta_0\sin^4(q_p/2)|B|^2B\vphantom{1}\\
+i\frac{3|v_g|}{2\omega_p\gamma\sin^2(q_p/2)}\left(4|B|^2\frac{\partial
B}{\partial\xi}+B^2\frac{\partial B^*}{\partial\xi}\right)
-\frac{81}{8\omega_p^2\gamma\sin^2(q_p/2)}|B|^4B.
\end{multline}
After applying one last set of rescaling transformations,
\begin{equation} \label{scaling}
\begin{split}
&\tau \rightarrow \frac{9}{32} \frac{1}{\omega_p^2 \eta_0^2 \gamma
  \sin^{10}\left({q_p/2}\right)} \tau, \qquad
\xi \rightarrow \frac38 \frac{|v_g|}{\omega_p \eta_0 \gamma
  \sin^6\left({q_p/2}\right)} \xi,\\
&|B|^2 \rightarrow \frac{16}{27} \omega_p^2 \eta_0 \gamma
\sin^6\left({q_p/2}\right) |B|^2,\qquad{\rm and\ }\qquad g
\rightarrow \frac{32}{9} \omega_p^2 \eta_0^2
  \sin^8\left({q_p/2}\right)g,
\end{split}
\end{equation}
we end up with an amplitude equation governed by a single parameter,
\begin{equation} \label{BampEq}
\frac{\partial B}{\partial\tau}=gB +
\frac{\partial^{2}B}{\partial\xi^{2}} + i\frac{2}{3}
\left(4|B|^{2}\frac{\partial B}{\partial\xi} +
  B^{2}\frac{\partial B^{*}}{\partial\xi}\right)-2|B|^{2}B - |B|^{4}B.
\end{equation}
This amplitude equation which captures the slow dynamics of the
coupled oscillators just above the onset of the parametric
oscillations, is our central result. It is a variant of the
CGLE~\cite{AL02}, supplemented with the uncommon nonlinear gradient
terms. Brand \emph{et. al.}~\cite{BRLONEW86,BRDE89,DEBR90} studied
similar equations that appear in the theory of binary fluid
mixtures. However in their work all the terms in Eq.~(\ref{BampEq})
had complex coefficients, yielding localized solutions on which the
influence of the nonlinear gradients was studied. Here we focus on
extended solutions and show that the nonlinear gradient terms yield
the wave number detunings of single mode solutions that, as far as
we are aware of, were previously introduced by hand.
\chapter{Single Mode Oscillations}
\section{Single Mode Solution of The Amplitude
Equation}\label{SMsolution}
Once we have obtained the amplitude equation~(\ref{BampEq}) it can
be used to study a variety of dynamical solutions, ranging from
simple single-mode to more complicated nonlinear extended solutions,
or possibly even localized solutions. Here we focus on the regime of
small reduced amplitude $g$ and look upon the saturation of
single-mode solutions of the form
\begin{equation}\label{SingleMode}
B=b_ke^{-ik\xi},\quad {\rm with}\quad b_k=|b_k|e^{i\varphi},
\end{equation}
corresponding to a standing wave, determined by Eq.~(\ref{u0withB})
to be
\begin{equation}
u_0=4|b_k|\cos(\omega_pt-\pi/4)\cos(qx-\pi/4-\varphi).
\end{equation}
The shifted wave number $q$ is given by
\begin{equation}
q=q_p + \frac83\frac{\omega_p\eta_0 \gamma
\sin^6\left({q_p/2}\right)}{|v_g|} \epsilon \sqrt{\delta}k.
\end{equation}
The original boundary conditions $u(0,t)=u(N+1,t)=0$ impose
$\varphi=\pi/4$ and require that the wave numbers $q$ be quantized
according to Eq.~(\ref{StandingWaves}).

We substitute (\ref{SingleMode}) into the amplitude equation
(\ref{BampEq}) and obtain
\begin{equation}\label{SMeq}
\frac{\partial b_k}{\partial\tau}=\left(g-k^2\right)b_k +
2(k-1)|b_k|^{2}b_k - |b_k|^{4}b_k.
\end{equation}
From the linear terms in Eq.~(\ref{SMeq}) we find, as expected, that
for $g>k^2$ the zero-displacement solution is unstable to small
perturbations of the form of (\ref{SingleMode}), defining the
parabolic neutral stability curve, shown as a dashed line in
Fig.~\ref{StabilityBal}. The nonlinear gradients and the cubic term
take the simple form $2(k-1)|b_k|^2b_k$. For $k < 1$ these terms
immediately act to saturate the growth of the amplitude assisted by
the quintic term. Standing waves therefore bifurcate supercritically
from the zero-displacement state. For $k > 1$ the cubic terms act to
increase the growth of the amplitude, and saturation is achieved
only by the quintic term. Standing waves therefore bifurcate
subcritically from the zero-displacement state. It should be noted
that similar wave-number dependent bifurcations were also predicted
and observed numerically in Faraday
waves~\cite{milner,peilong00,peilong02}, though in these works the
wave number detuning $k$ was introduced \emph{a priori} and was not
obtained directly from the amplitude equations. Here the detuning is
a direct consequence of the nonlinear gradient terms in
Eq.~(\ref{BampEq}), and is explicitly related to the physical
parameters of the system.

\begin{figure}
\begin{center}
\includegraphics[width=0.75\columnwidth]{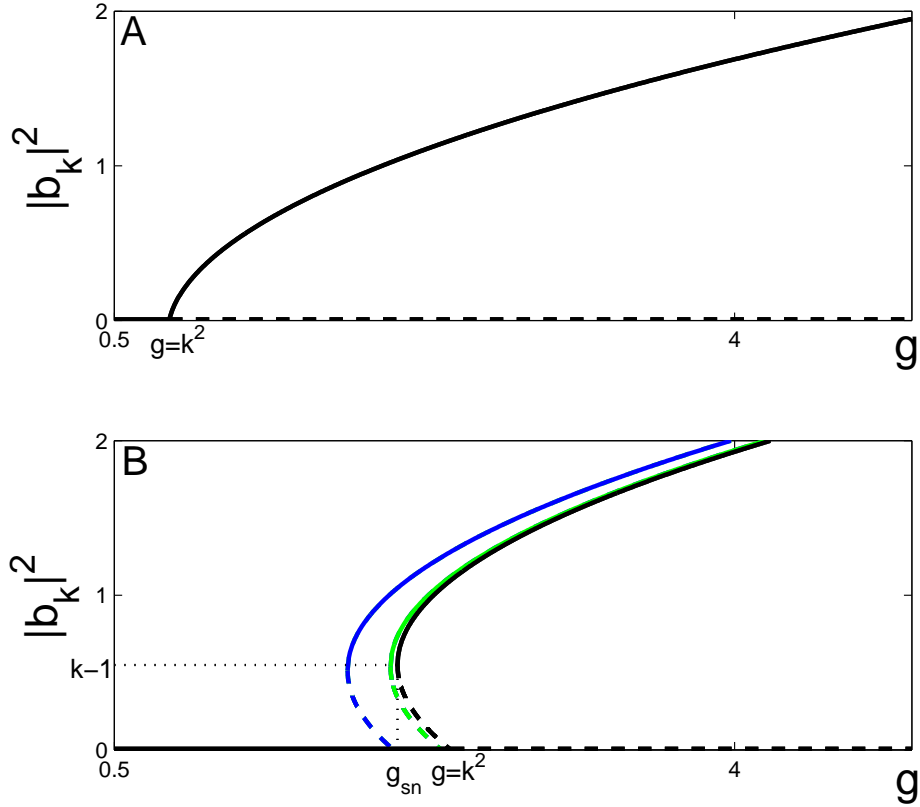}%
\caption{\label{fig:AnalyticBif}The amplitude of oscillations of the
single mode state $|b_k|^2$ plotted as a function of the reduced
driving amplitude g for two values of $k$. Solid and dashed black
lines are the positive and negative square root branches of the
calculated response  in~(\ref{Bsteady}), the latter clearly
unstable. (A) $k=0.9$ yields a supercritical bifurcation from the
zero displacement state into the single mode state with a
continuously increasing amplitude. (B) For $k=1.55$ a subcritical
bifurcation occurs at $g=k^2$. Colored lines mark the single mode
solutions calculated by Eq.~(\ref{SMc}) for $\epsilon=0.01$ (blue
lines) and $\epsilon=0.001$ (green line). Both solutions are
calculated for $\delta=0.01$ and the rest of the parameters are
chosen to yield $k=1.55$. A good agreement with Eq.~(\ref{Bsteady})
is verified for $\epsilon\ll\sqrt\delta$.}
\end{center}
\end{figure}

The saturated amplitude $|b_k|$, obtained by setting
Eq.~(\ref{SMeq}) to zero, is given by
\begin{equation}\label{Bsteady}
|b_k|^2 = (k-1) \pm \sqrt{(k-1)^2 + (g-k^2)}.
\end{equation}
In Fig.~\ref{fig:AnalyticBif} we plot $|b_k|^2$ as a function of $g$
for two values of $k$. The solid (dashed) lines are stable
(unstable) states. For $k<1$ only the positive square root branch of
Eq.~(\ref{Bsteady}) is obtained, and the amplitude of oscillations
increases continuously from zero for $g>k^2$. A subcritical
bifurcation is obtained for $k>1$, the negative square root branch
is obtained in addition to the positive branch for reduced driving
amplitudes ranging from $k^2$ down to the saddle-node point given by
\begin{equation}\label{SN}
  g_{sn}=2k-1.
\end{equation}
Since two stable solutions are obtained in this range of reduced
driving amplitudes for subcritical bifurcations, the system exhibits
hysteresis for quasistatic driving amplitude sweeps. Quasistatic
sweeps mean that the change in the driving amplitude is much slower
than the time required for transients to fade so that a steady-state
is obtained between each change of the driving amplitude.
\section{Comparison With The Exact Form of Single Mode Solutions}
We now wish to compare the exact form of single mode solutions
(\ref{SMSolution}), obtained  by Lifshitz and Cross, with the
solutions obtained in the previous section. In the limit of driving
amplitudes just above threshold and $\eta\ll1$, the exact amplitudes
of oscillations (\ref{SMc}) correspond to Eq.~(\ref{Bsteady}). The
excited mode $m$ is the mode which minimizes $h_m(\omega_p)$, and
its wave number $q_m$ and frequency $\omega_m$ are identified (up to
$O(\frac1N)$ corrections) with $q_p$ and $\omega_p$ respectively.
The frequency detuning $\Omega_m$ corresponds to
$\epsilon\sqrt\delta v_g k$, with $\Omega_c$ equivalent to $k=1$.
This implies that for a truly continuous extended system the
standing waves will always bifurcate supercritically with a wave
number $q_p$ as long as $g$ is increased quasistatically from zero.
It is the discreteness of the normal modes which provides the
detuning parameter essential for a subcritical bifurcation for a
quasistatic increase of the driving amplitude. In
Fig.~\ref{fig:AnalyticBif} (B) we compare two exact single mode
solutions calculated by Eq.~(\ref{SMc}) with the solution of
Eq.~(\ref{Bsteady}) for $k=1.55$. It is demonstrated that the
agreement between the amplitude equation approach solution and the
exact solution is better for smaller $\epsilon/\sqrt\delta$, as
discussed in section~\ref{LinearStabiltyAeq}.

For frequency detunings and amplitudes of oscillations of order
$\epsilon\sqrt\delta$, the phase obtained from Eq.~(\ref{SMb}) is
$\varphi=\pi/4+O(\epsilon\sqrt\delta)$, in agreement with the phase
obtained from the amplitude equation.

\section{Secondary Instabilities}\label{sndins}
\begin{figure}
\begin{center}
\includegraphics[width=0.75\columnwidth]{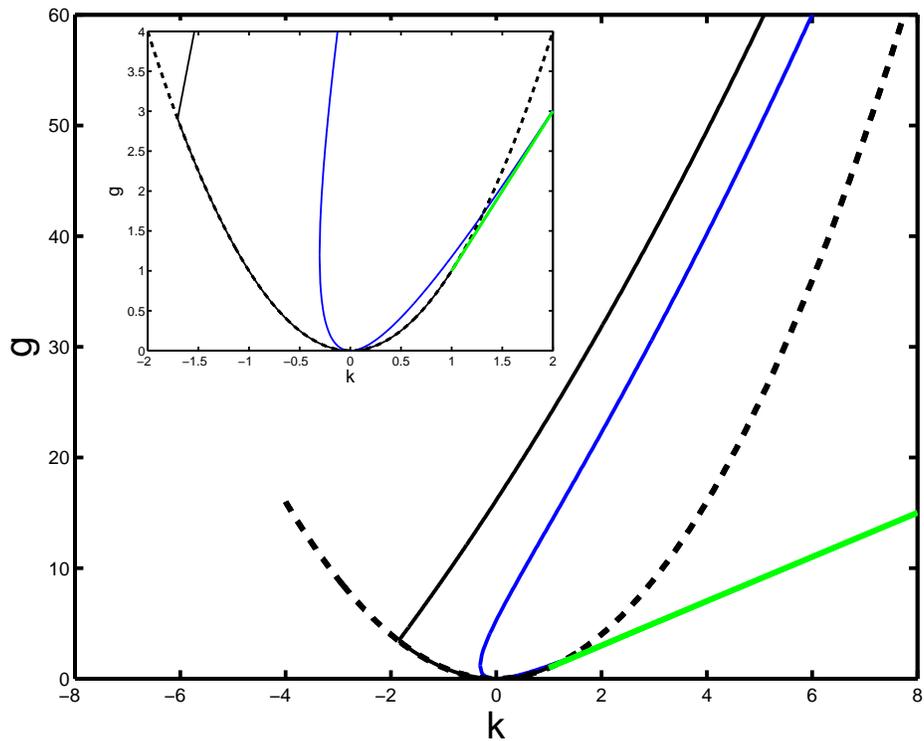}%
\caption{\label{StabilityBal}
  Stability boundaries of the single-mode solution of
  the amplitude equation~(\ref{BampEq}) in the $g$~{\it vs.}~$k$ plane. Dashed line:
  neutral stability boundary $g=k^2$. Solid lines: stability boundary of the
  single-mode solution~(\ref{SingleMode}) as obtained by Eqs.~(\ref{stabcurves}) for a continuous spectrum (blue curve)
  and for $N=92$ with the same parameters given in section~\ref{sec71} (black
  curve). Green line: saddle node point for the subcritical bifurcation
  $g_{sn}=2k-1$ which coincides with the discrete stability boundary
  for $k>1$.}
\end{center}
\end{figure}

We study secondary instabilities of the single mode solutions by
performing a linear stability analysis of the solution
(\ref{SingleMode}). We substitute
\begin{equation}\label{BStability}
B=b_ke^{-ik\xi}+\left(\beta_+
e^{-i(k+Q)\xi}+\beta_-^*e^{-i(k-Q)\xi}\right),
\end{equation}
with $|\beta_\pm|\ll1$, into the amplitude equation~(\ref{BampEq})
and linearize in $\beta_\pm$. Since the amplitude
equation~(\ref{BampEq}) is invariant under phase transformations
$B\rightarrow Be^{-i\varphi}$, the stability of the single mode
solution cannot depend on the phase of $b_k$. We therefore assume
that $b_k$ is real, and linearize the following terms of
Eq.~(\ref{BampEq})
\begin{eqnarray}
|B|^{2}&\rightarrow&b_{k}^{2}+b_{k}\left((\beta_{+}+\beta_{-})e^{-iQ\xi}+(\beta_{+}^{*}+\beta_{-}^{*})e^{iQ\xi}\right)\,;\nonumber\\
B^{2}&\rightarrow&e^{-i2k\xi}\left(b_{k}^{2}+b_{k}\left(2\beta_{+}e^{-iQ\xi}+2\beta_{-}^{*}e^{iQ\xi}\right)\right)\nonumber\,;\\
|B|^{4}&\rightarrow&b_{k}^{4}+b_{k}^{3}\left(2(\beta_{+}+\beta_{-})e^{-iQ\xi}+2(\beta_{+}^{*}+\beta_{-}^{*})e^{iQ\xi}\right)\nonumber\,;\\
|B|^{2}B&\rightarrow&e^{-ik\xi}\left(b_{k}^{3}+b_{k}^{2}\left((2\beta_{+}+\beta_{-})e^{-iQ\xi}+(\beta_{+}^{*}+2\beta_{-}^{*})e^{iQ\xi}\right)\right)\nonumber\,;\\
|B|^{4}B&\rightarrow&e^{-ik\xi}\left(b_{k}^{5}+b_{k}^{4}\left((3\beta_{+}+2\beta_{-})e^{-iQ\xi}+(2\beta_{+}^{*}+3\beta_{-}^{*})e^{iQ\xi}\right)\right)\nonumber\,;\\
\frac{\partial B}{\partial\xi}&\rightarrow&
e^{-ik\xi}\left(-ikb_{k}-i(k+Q)\beta_{+}e^{-iQ\xi}-i(k-Q)\beta_{-}^{*}e^{iQ\xi}\right)\nonumber\,;\\
|B|^2\frac{\partial
B}{\partial\xi}&\rightarrow&e^{-ik\xi}\left(-ikb_{k}^{3}-ib_{k}^{2}\left(((2k+Q)\beta_{+}+k\beta_{-})e^{-iQ\xi}+
(k\beta_{+}^{*}+(2k-Q)\beta_{-}^{*})e^{iQ\xi}\right)\right)\nonumber\,;\\
\frac{\partial B^{*}}{\partial\xi}&\rightarrow&
e^{ik\xi}\left(ikb_{k}+i(k+Q)\beta_{+}^{*}e^{iQ\xi}+i(k-Q)\beta_{-}e^{-iQ\xi}\right)\nonumber\,;\\
B^{2}\frac{\partial
B^{*}}{\partial\xi}&\rightarrow&e^{-ik\xi}\left(ikb_{k}^{3}+ib_{k}^{2}\left((2k\beta_{+}+(k-Q)\beta_{-})e^{-iQ\xi}+
((k+Q)\beta_{+}^{*}+2k\beta_{-}^{*})e^{iQ\xi}\right)\right).\nonumber\\
\end{eqnarray}
The terms of order one of the equation obtained from the
linearization of Eq.~(\ref{BampEq}) recover the same
Eq.~(\ref{SMeq}) for $b_k$. The terms with spatial dependence of
$e^{-iQ\xi}$ must satisfy
\begin{equation}\label{betaplus}
\begin{split}
\frac{\partial\beta_+}{\partial\tau}&=g\beta_+-(k+Q)^2\beta_++\frac23b_k^2
\left(\vphantom{{1^1}^1}4\left((2k+Q)\beta_{+}+k\beta_{-}\right)-2k\beta_{+}-(k-Q)\beta_{-}\right)\\
&\quad-2(2\beta_{+}+\beta_{-})b_k^2-(3\beta_{+}+2\beta_{-})b_k^4\\
&=\left[\vphantom{{1^1}^1}g-(k+Q)^2+b_k^2(4(k-1)-3b_k^2+8Q/3)\right]\beta_+
+2b_k^2\left[\vphantom{{1^1}^1}k-1-b_k^2+Q/3\right]\beta_-.
\end{split}
\end{equation}
The terms with spatial dependence of $e^{iQ\xi}$ similarly obey
\begin{equation}\label{betaminus}
\begin{split}
  \frac{\partial\beta_-}{\partial\tau}&=g\beta_--(k-Q)^2\beta_-+\frac23b_k^2
  \left(\vphantom{{1^1}^1}4(k\beta_{+}+(2k-Q)\beta_{-})-(k+Q)\beta_{+}-2k\beta_{-}\right)\\
  &\quad-2(\beta_{+}+2\beta_{-})b_k^2-(2\beta_{+}+3\beta_{-})b_k^4\\
  &=\left[\vphantom{{1^1}^1}g-(k-Q)^2+b_k^2(4(k-1)-3b_k^2-8Q/3)\right]\beta_-+
  2b_k^2\left[\vphantom{{1^1}^1}k-1-b_k^2-Q/3\right]\beta_+.
\end{split}
\end{equation}
Using Eq.~(\ref{Bsteady}) we find that (for $\partial
b_k/\partial\tau=0$)
\begin{equation}
\begin{split}
g-k^2+4(k-1)b_k^2-3b_k^4=
2b_k^2(k-1-b_k^2),
\end{split}
\end{equation}
and write Eqs.~(\ref{betaplus}) and (\ref{betaminus}) in matrix form
as
\begin{eqnarray}
&&\frac\partial{\partial\tau}\left(
\begin{array}{c}
      \beta_+ \\
      \beta_- \\
\end{array}\right)=M\left(
\begin{array}{c}
      \beta_+ \\
      \beta_- \\
\end{array}
\right),\quad{\rm where}\\&& M\equiv
\left(%
\begin{array}{cc}
 2b_k^2(k-1-b_k^2)-Q^2-2Q(k-4b_k^2/3) & 2b_k^2(k-1-b_k^2+Q/3) \\
 2b_k^2(k-1-b_k^2-Q/3) & 2b_k^2(k-1-b_k^2)-Q^2+2Q(k-4b_k^2/3) \\
\end{array}%
\right).\nonumber
\end{eqnarray}
We express the eigenvalues of the matrix $M$ using its trace $trM$
and its determinant |M|,
\begin{equation}\label{betagrowthrates}
  \sigma_\pm=\frac{trM}{2}\pm\sqrt{\left(\frac{trM}{2}\right)^2-|M|}.
\end{equation}
The single mode solution (\ref{SingleMode}) is a stable solution
only if all wave numbers $Q$ yield negative eigenvalues for $M$,
obtained for $trM<0$ and $|M|>0$. A negative trace of the matrix $M$
is obtained for all wave numbers $Q$ if
\begin{equation}\label{minusbranchstb}
b_k^2>k-1\qquad{\rm and}\qquad b_k^2>0.
\end{equation}
Thus the negative square root branch in~(\ref{Bsteady}) obtained for
subcritical bifurcations $k>1$, is confirmed to be always unstable.
The stability of the positive square root is determined by the
constraint on the determinant of $M$,
\begin{eqnarray}\label{PosDet}
|M|&=&\left(2b_k^2(k-1-b_k^2)-Q^2\right)^2-4Q^2(k-4b_k^2/3)^2
-4b_k^4(k-1-b_k^2)^2+4b_k^4Q^2/9\nonumber\\
&=&Q^4-4Q^2b_k^2(k-1-b_k^2)-4Q^2(k^2-8b_k^2k/3+16b_k^4/9)+4b_k^4Q^2/9\nonumber\\
&=&\frac83Q^2\left(\frac38Q^2-b_k^4+\frac{5k+3}{2}b_k^2-\frac32k^2\right)>0.
\end{eqnarray}
In order to obtain stable single modes, this inequality should be
satisfied for all wave numbers $Q>0$\;\footnote{Perturbations with
the same wave number as of the single mode, corresponding to $Q=0$,
decay as long as inequality~(\ref{minusbranchstb}) holds. This
confirms the stability of the single mode solution towards
perturbations with the same wave number mentioned in
chapter~\ref{SMresponse}. Since $Q=0$ is a degenerate case of the
ansatz~(\ref{BStability}), we can take $\beta_-=0$ and consider only
the sign of $k-1-b_k^2$, as apparent from Eq.~(\ref{betaplus}). The
single mode solution is therefore stable for the positive square
root branch of Eq.~~(\ref{Bsteady}) ($b_k^2>k-1$), and unstable for
the negative branch ($b_k^2<k-1$).}. Thus the
inequality~(\ref{PosDet}) can be replaced with
\begin{equation}
  b_k^4-\frac{5k+3}{2}b_k^2+\frac32k^2 <\frac38Q_{min}^2,
\end{equation}
where
\begin{equation}\label{Qmin}
  Q_{min}\equiv\frac{3|v_g|}{8\omega_p\eta_0\gamma\sin^6(q_p/2)\epsilon\sqrt\delta}\frac\pi{N+1}
\end{equation}
is the mode to mode separation in our rescaled units. Stable single
mode oscillations therefore must have amplitudes obeying

\begin{subequations}\label{stabcurves}
\begin{align}
  &b_k^2>0\quad{\rm for}\quad k<1,\label{stabcurves_a}\\
  &b_k^2>k-1\quad{\rm for}\quad k>1,\label{stabcurves_b}\\\displaybreak[0]
  &b_k^2>\frac{5k+3}{4}-\sqrt{\left(\frac{5k+3}{4}\right)^2-
   \frac32k^2+\frac38Q_{min}^2},\quad{\rm and}\label{stabcurves_c}\\
  &b_k^2<\frac{5k+3}{4}+\sqrt{\left(\frac{5k+3}{4}\right)^2-
   \frac32k^2+\frac38Q_{min}^2}\label{stabcurves_d}.
\end{align}
\end{subequations}
The right hand side of inequality~(\ref{stabcurves_d}) bounds the
amplitude of stable single mode oscillations from above. The lower
bound of the stable amplitudes is determined by either one of the
inequalities (\ref{stabcurves_a})-(\ref{stabcurves_c}), depending on
values of $k$ and $Q_{min}$. For $k<1$ the right hand side of
inequality~(\ref{stabcurves_c}) determines the lower bound if
\begin{eqnarray}
  &&\frac{5k+3}{4}-\sqrt{\left(\frac{5k+3}{4}\right)^2-
   \frac32k^2+\frac38Q_{min}^2}>0\nonumber\\
  &&\Leftrightarrow\frac32k^2>\frac38Q_{min}^2\nonumber\\
  &&\Leftrightarrow|k|>\frac{Q_{min}}2,
\end{eqnarray}
while for  $k>1$ the condition states
\begin{eqnarray}
  &&\frac{5k+3}{4}-\sqrt{\left(\frac{5k+3}{4}\right)^2-
  \frac32k^2+\frac38Q_{min}^2}>k-1\nonumber\\
  &&\Leftrightarrow(k-1)^2-\frac12(5k+3)(k-1)+\frac32k^2-\frac38Q_{min}^2>0 \nonumber\\
  &&\Leftrightarrow-(k-1)\left(\frac32k+\frac52\right)+\frac32k^2- \frac38Q_{min}^2>0\nonumber\\
  &&\Leftrightarrow1<k<\frac52-\frac32\left(\frac{Q_{min}}2\right)^2.\nonumber\\
\end{eqnarray}
In the experimental scheme we focus on (see
chapter~\ref{chp:numerics}) the bifurcation from the zero
displacement state to single mode oscillations always occur for
$|k|<Q_{min}/2$. The stability of the single mode oscillations is
therefore bounded from below by (\ref{stabcurves_a}) and
(\ref{stabcurves_b}). Using Eq.~(\ref{Bsteady}) we can write the
boundaries of stable single mode oscillations as a function of the
reduced driving amplitude $g$ and $k$. The lower bounds determined
by inequalities~(\ref{stabcurves_a}) and (\ref{stabcurves_b}) then
take the simple form
\begin{equation}\label{lowerbound}
g=k^2\quad{\rm and}\quad g=2k-1
\end{equation}
respectively. We thus obtain that for a supercritical (subcritical)
bifurcation, the positive square root branch of Eq.~(\ref{Bsteady})
is stable for reduced driving amplitudes $g$ just above the  neutral
stability (saddle-node) point. In Fig.~\ref{StabilityBal} we show
the stability boundaries of single mode oscillations in the $(g,k)$
plane, describing the so-called \emph{stability
balloon}~\cite{reviewcross} of the single mode state. The solid
black curve is the stability balloon obtained for a particular set
of parameters, such that $Q_{min}/2>1$, and thus it is bounded from
bellow by Eqs.~(\ref{lowerbound}). Outside the curve the
oscillations undergo instabilities with $Q=Q_{min}$. In the limit of
very large arrays $Q_{min}\rightarrow0$, thus the single-mode state
is unstable towards perturbations with $Q\rightarrow0$, typical of
Eckhaus~\cite{eckhaus} instabilities in extended systems. In this
limit the stability balloon shrinks to the blue curve in
Fig.\ref{StabilityBal}.

\newpage

\chapter{Numerical Simulations of Possible Experiments}
\label{chp:numerics} The purpose of our numerical work is twofold:
to test our analytical predictions and to simulate possible
experiments that may verify our results. The equations of
motion~(\ref{eom}) are integrated numerically using the fourth-order
Runge-Kutta method. Most of our numerical simulations mimic an
experiment in which the ac component $H$ is increased
quasistatically from zero, until moderate oscillations are obtained,
and then swept back to zero. This is done by fixing the parameters
$\Gamma,\;\Delta,\;\omega_p,\;\eta$ and the number of oscillators in
the array $N$, and scanning the values of $H$.
Starting with driving amplitudes lower than the threshold
$2\epsilon\gamma\omega_p/\Delta^2$, the initial conditions are taken
to be those of the zero-displacement state $u_n(t=0)=\dot
u_n(t=0)=0$ for all $n$, superimposed with small random noise. The
displacements of the beams $u_n(t)$ develop in time according to the
equations of motion~(\ref{eom}). Once a steady state is obtained,
the time average of the squared amplitude of each beam is
calculated. The driving amplitude $H$ is then increased, and the
equations are integrated for the new $H$, using the displacements
and velocities of the beams just before $H$ was increased as the new
initial conditions. Again a small random noise is added to the
initial conditions in order to prevent the system from staying on an
unstable solution of the equations, solutions which are not
accessible in experiment. By repeating this procedure for increasing
drives, the amplitude of oscillations as a function of the driving
amplitude is obtained, and by switching to decreasing sweeps of $H$
after the array is excited, the hysteretic character of subcritical
bifurcations can be observed.

We test numerically our two main predictions of the amplitude
equation~(\ref{BampEq}): the existence of a wave-number dependent
bifurcation and the stability boundaries of single mode
oscillations.

\section{Wave-Number Dependent Bifurcation}\label{sec71}
The first mode to emerge when increasing the driving amplitude from
$H=0$ is the mode which minimizes Eq.~(\ref{hcm}), whose wave number
$q_m$ is the closest to $q_p$ among the spectrum of vibrational
modes. In order to obtain the influence of the wave number detuning
$k$ on the type of bifurcation from a zero-displacement state to
single-mode oscillations, the experimenter must have a way to
control $q_m$. By changing the number of oscillators $N$ in the
array, the spectrum of vibrational modes can be modified, yielding
different wave number detunings for the same driving frequency
$\omega_p$. Since $q_m$ and $q_p$ can differ by up to $\pi/(N+1)$,
$|k|$ is bounded from above by
\begin{equation}
  |k|<\frac{Q_{min}}2,
\end{equation}
where $Q_{min}$ is given by Eq.~(\ref{Qmin}). In the limit of vary
large arrays $N\rightarrow\infty$, $q_m\rightarrow q_p$ and the
bifurcations are always supercritical\footnote{More accurately $q_m
\rightarrow q_p+O(\epsilon^2)$, as explained in
section~\ref{Aderivation}}. Thus if one wants to prevent hysteresis
effects in an experiment, large arrays should be considered. The
weaker the dissipation of the system, the larger the array should
be. For smaller arrays such that $Q_{min}\gtrsim 2$, either
supercritical bifurcations $(k<1)$ or subcritical bifurcations
$(k>1)$ can be obtained, depending on the specific values of
$\omega_p$ and $\Delta$.

Our analytical treatment is valid for $\epsilon,\delta\ll1$ and
$N\gg1$. We therefore choose to integrate arrays of about $100$
oscillators with dissipation coefficients $\Gamma=0.01$ and
$\eta=0.1$, corresponding to $\epsilon=0.01$ and $\delta=0.01$
respectively. Setting the dc component to $\Delta^2=0.25$, ensures
that the spectrum of vibrational modes satisfy $Q_{min}\simeq2$, and
thus both types of bifurcations can be obtained, depending on the
number of oscillators in the array. We choose the driving frequency
\begin{equation}
\omega_p=\sqrt{1-2\Delta^2\sin^2(q_ {73}/2)}=0.767445\ldots,
\end{equation}
to obtain $q_m=q_p$ for an array of $N=100$ oscillators and a mode
number $m=73$.

\begin{figure}
\begin{center}
\includegraphics[width=1\columnwidth]{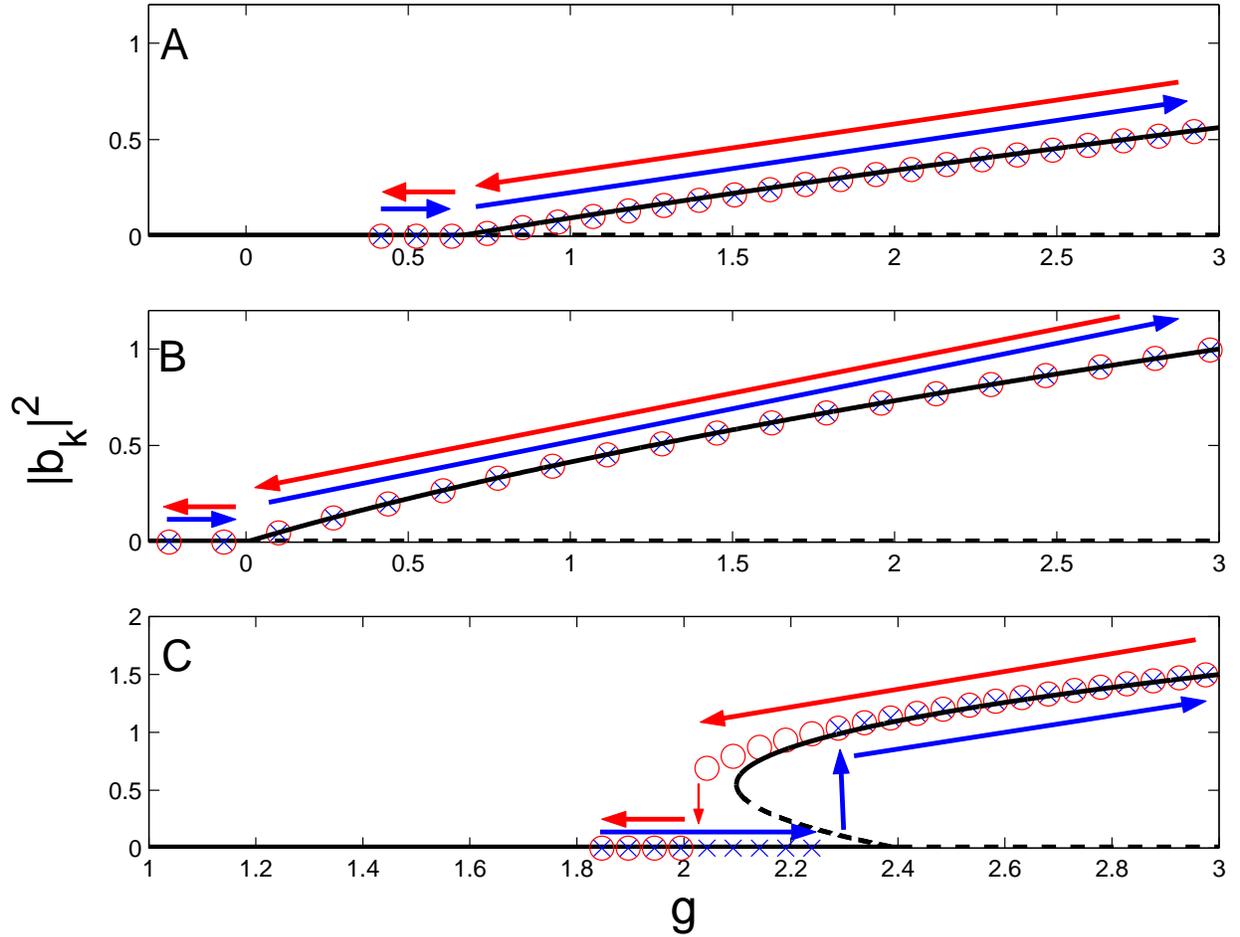}%
\caption{\label{NumericsFig} Response of the oscillator
  array plotted as a function of reduced amplitude $g$ for three
  different wave number shifts determined by fixing the number of oscillators
  $N$. (A) $k\simeq-0.81$ obtained for $N=92$, (B) $k=0$ obtained for $N=100$, and (C) $k\simeq1.55$ obtained for $N=98$. Solid and dashed lines are the positive and
  negative square root branches of the calculated response
  in~(\ref{Bsteady}), the latter clearly unstable. The symbols are
  numerical values obtained by numerical integration of the equations of
  motion~(\ref{eom}). Blue x-marks indicate solutions obtained for increasing sweeps of $g$,
  while red circles are for decreasing sweeps. The curves in (A) and (B)
  show a  supercritical bifurcation, while that of (C) exhibits a subcritical bifurcation with clear hysteresis.
  The values of the parameters of the array are
  $\Delta=0.5$, $\omega_p=0.767445$, $\epsilon\gamma=0.01$ and $\eta=0.1$.}
\end{center}
\end{figure}

We numerically integrate the equations of motion~(\ref{eom}) with
the above parameters for arrays with $N=100$, $N=98$ and $N=92$,
corresponding to wave number detunings of $k=0$, $k\simeq-0.81$ and
$k\simeq1.55$ respectively. For each array the scenario of
increasing sweeps of the drive $H$ followed by decreasing sweeps as
described above is performed. For each driving amplitude, the
Fourier components of the steady-state solution are computed to
verify that only single modes are found, suggesting that in this
regime of parameters only these states are stable. In
Fig.~\ref{NumericsFig} we plot $|b_k|^2$ as a function of the
reduced driving amplitude $g$ for the three wave number shifts $k$.
The symbols are the numerically computed Fourier components, blue x
marks are calculated for increasing driving amplitude sweeps and red
circles are calculated for decreasing sweeps, showing clear
hysteresis for $k>1$. The solid (dashed) lines are the stable
(unstable) solutions of Eq.~(\ref{Bsteady}). The disagreement
between the analytic curves and the numeric integration lies within
the scope of accuracy that the amplitude equations approach as
discussed previously ($g$ for instance is determined up to
$O(\frac{\epsilon}{\sqrt\delta})\sim0.1$).

\begin{figure}
\begin{center}
\includegraphics[width=1\columnwidth]{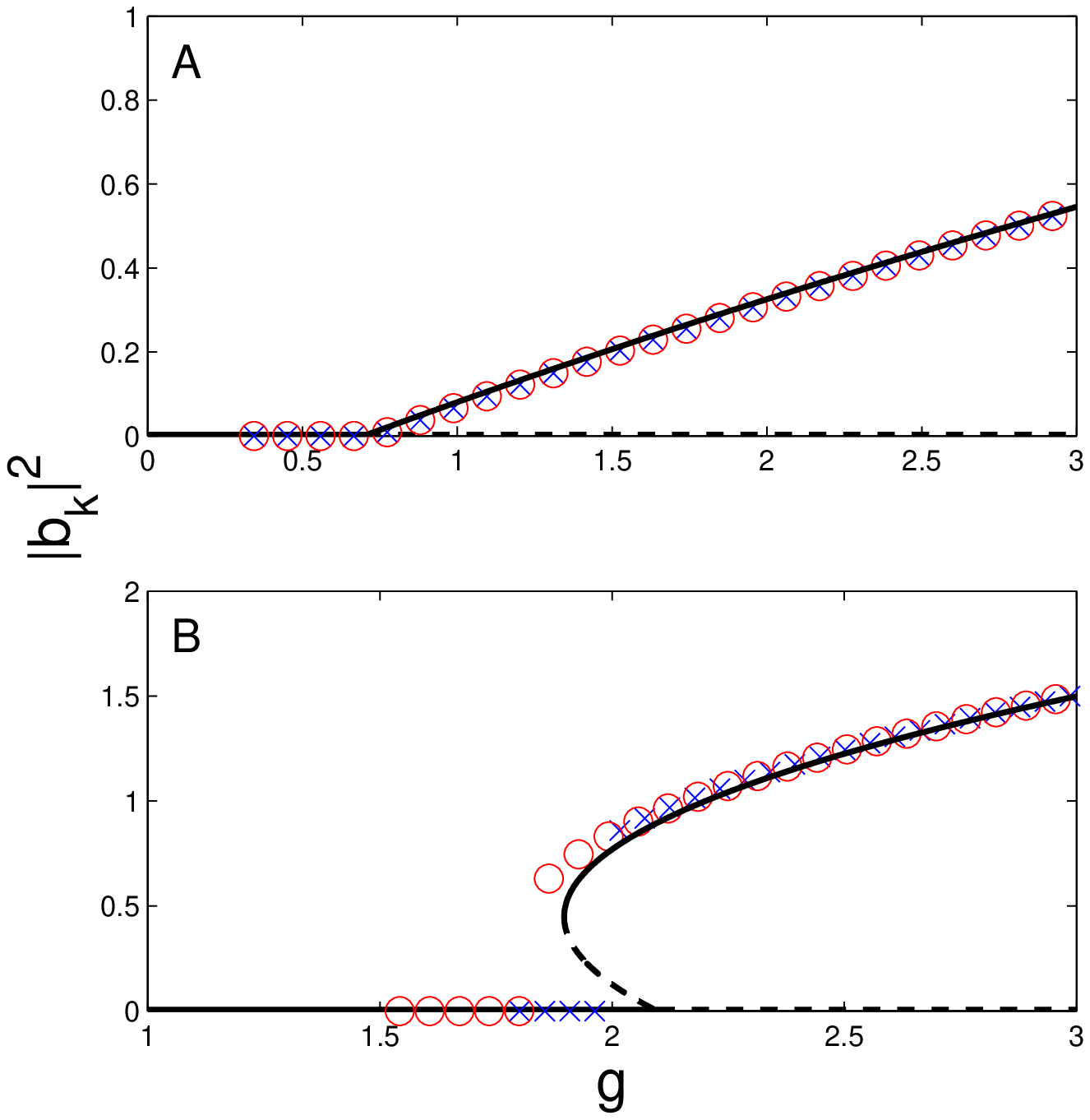}%
\caption{\label{NumerDelta} Response of an
  array of $N=100$ oscillators plotted as a function of reduced amplitude $g$ for two wave-number shifts
  determined by changing the dc component of the potential difference between the beams.
  (A) $k\simeq-0.84$ obtained for $\Delta=0.4991$ bifurcates
  supercritically, and (B) $k\simeq1.44$ obtained for $\Delta=0.5015$ bifurcates subcritically.
  All other parameters are set to the same values given in Fig.~\ref{NumericsFig}.}
\end{center}
\end{figure}

In experiment it might be easier to control $k$ by changing the dc
component of the potential difference between the beams, thus
changing $q_p$ rather than $q_m$. In Fig.~\ref{NumerDelta} we plot
the response of an array of $N=100$ oscillators for two different dc
components, $\Delta=0.491$ and $\Delta=0.5015$. The values of all
other parameters are as given above. Changing the dc component of
the potential difference might however change other parameters of
the array, such as the dissipation coefficients (see
chapter~\ref{BRexperiment}) or the coefficient of the cubic elastic
restoring force~\cite{turner02}. In addition, a change of $q_p$
involves a change in the scalings performed in our analytic
treatment. Thus, for the sake of simplicity we have concentrated on
controlling $k$ by changing the number of oscillators $N$, but in
experiment this will be difficult to do.

\section{Secondary Bifurcations}
\begin{figure}
\begin{center}
\includegraphics[width=0.75\columnwidth]{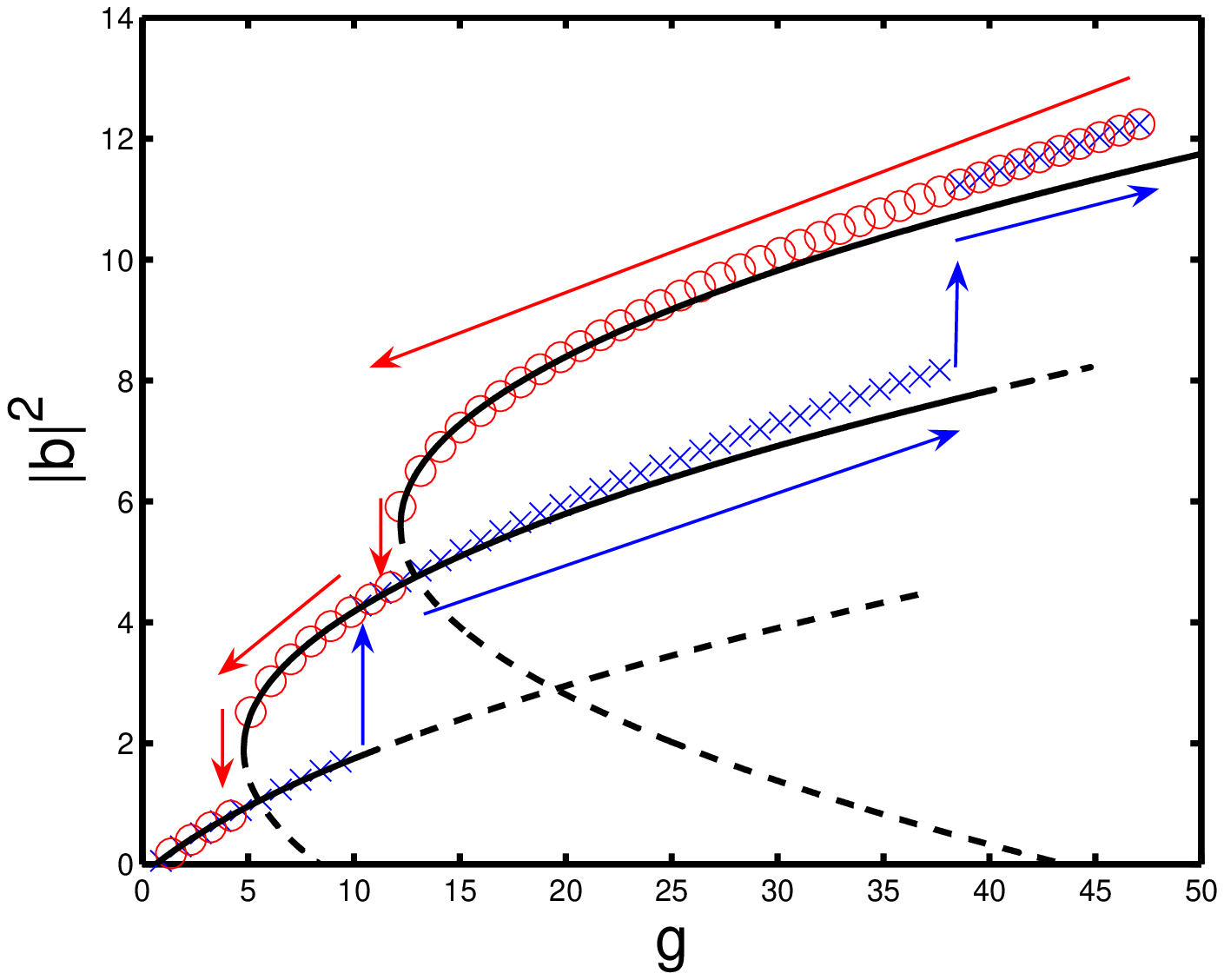}%
\caption{\label{fig:SndIns}
  A sequence of secondary instabilities following the
  initial onset of single-mode oscillations in an array of $92$ beams
  and the same parameters given in Fig.~\ref{NumericsFig} plotted
  as a function of the reduced driving amplitude $g$. Solid (dashed) lines are stable (unstable)
  solutions defined by (\ref{Bsteady}), for $k=-0.81$, $k=2.90$ and $k=6.60$
  corresponding to the first wave number to emerge and two shifts of the wave number by $Q_{min}$ and
  $2Q_{min}$ respectively. Numerical integration of the equations of
  motion~(\ref{eom}) for an upward sweep of $g$ (blue
  x-marks), followed by a downward sweep (red circles) exhibits a
  strong hysteresis and confirms the theoretical predictions for the
  stability of the single mode oscillations
  as illustrated in Fig.~\ref{StbBal:SndIns}.}
\end{center}
\end{figure}

\begin{figure}
\begin{center}
\includegraphics[width=0.75\columnwidth]{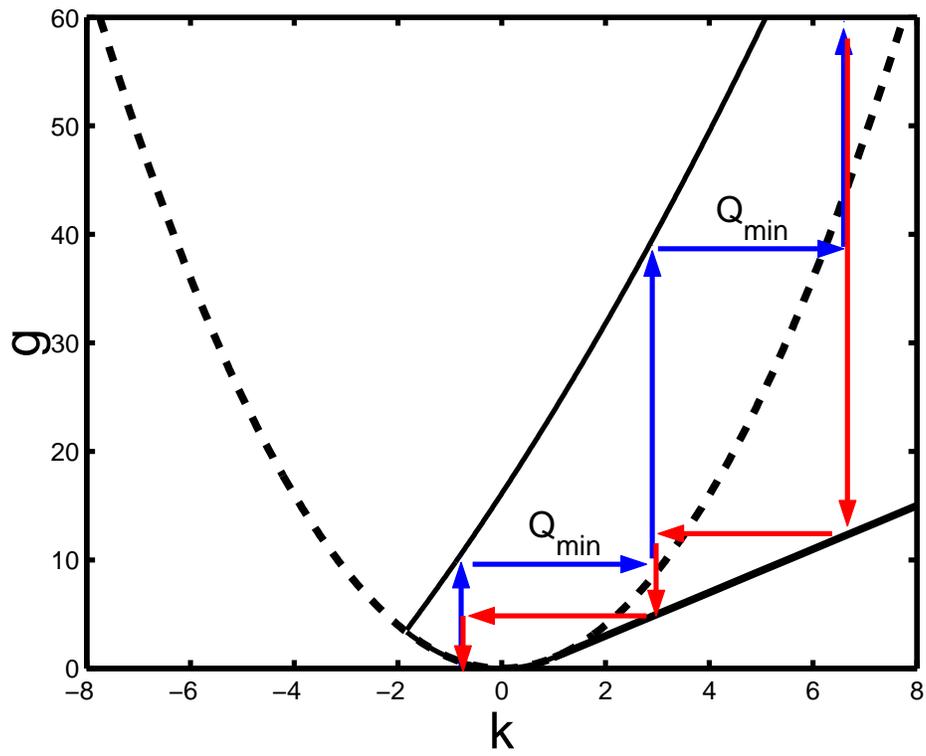}%
\caption{\label{StbBal:SndIns}
  The stability balloon given in Fig.~\ref{StabilityBal} superimposed with vertical and
  horizontal arrows that mark the secondary instability transitions shown
  in Fig.~\ref{fig:SndIns}. Each transition shifts the wave-number detuning $k$ by $Q_{min}$.}
\end{center}
\end{figure}

The linear stability analysis of the standing wave state performed
in section~\ref{sndins} predicts a transition to a new standing wave
with a wave-number shift of a multiple of $\pi/(N+1)$ once the
driving amplitude is increased and has crossed the upper bound of
the stability balloon. Since the upper bound monotonically increases
with $k$, the new wave number will always be larger. A sequence of
three transitions, obtained numerically, is shown in
Fig.~\ref{fig:SndIns}, superimposed with our theoretical
predictions. The sequence of transitions is also sketched for
comparison within the stability balloon in Fig.~\ref{StbBal:SndIns}.
All wave-number shifts obtained numerically are of $\pi/(N+1)$.
Secondary instabilities are therefore another scenario for the
experimental observation of hysteresis as a function of the applied
driving amplitude. Once a transition has occurred, the system will
return to its original state only when reducing the driving
amplitude below the saddle node point.

\chapter*{Conclusions} \label{conc}
\addcontentsline{toc}{chapter}{Conclusions} We derived amplitude
equations~(\ref{AmpEqs}) describing the response of large arrays of
nonlinear coupled oscillators to parametric excitation, directly
from the equations of motion yielding exact expressions for all the
coefficients. The dynamics at the onset of oscillations was studied
by reducing these two coupled equations into a single scaled
equation  governed by a single control parameter. Single mode
standing waves were found to be the initial states that develop just
above threshold, typical of parametric excitation. The single mode
oscillations bifurcate from the zero-displacement state either
supercritically or subcritically, depending on the wave number of
the oscillations. The wave number dependence originates in the
nonlinear gradient terms of the amplitude equation, which were
usually disregarded in the past by others in the analysis of
parametric oscillations above threshold. We also examined the
stability of single mode oscillations, predicting a transition of
the initial standing wave state to a new standing wave with a larger
wave number as the driving amplitude is increased.

In this work we showed that interesting response of coupled
nonlinear oscillators excited parametrically can also be obtained
for quasistatic driving amplitude sweeps, rather than frequency
sweeps usually preformed in experiments.  We proposed and
numerically demonstrated two experimental schemes for observing our
predictions, hoping to draw more attention of experimenters to the
dynamics produced by driving amplitude sweeps.

The results obtained by the numerical integration of the equations
of motion agree with our analysis, supporting the validity of the
amplitude equation~(\ref{BampEq}). We therefore believe that the
amplitude equations we derived can serve as a good starting point
for studying other possible states of the system. One particular
interesting dynamical behavior that can be considered is that of
localized modes, often observed in arrays of coupled nonlinear
oscillators and in other nonlinear systems as well. The conditions
for obtaining such modes and their dynamical properties could be
studied by looking for localized states of the amplitude equations.
Another interesting aspect that can be addressed using the amplitude
equations we derived is the response of the array to fast (rather
than quasistatic) driving amplitude sweeps, which should lead to
more complicated nonlinear response as observed by LC in their work.

\addcontentsline{toc}{chapter}{Bibliography}
\bibliography{ThesisBib}

\end{document}